\documentclass[sigconf, nonacm]{acmart}

\usepackage{graphicx}
\usepackage{textcomp}
\usepackage{xcolor}
\usepackage{algorithm}
\usepackage{algorithmic}
\usepackage{listings}
\usepackage{xcolor}
\usepackage{booktabs}
\usepackage{multirow}
\usepackage[most]{tcolorbox}

\definecolor{codegreen}{rgb}{0,0.6,0}
\definecolor{codegray}{rgb}{0.5,0.5,0.5}
\definecolor{codepurple}{rgb}{0.58,0,0.82}
\definecolor{backcolour}{rgb}{0.95,0.95,0.92}

\lstdefinestyle{codestyle}{
  backgroundcolor=\color{backcolour}, commentstyle=\color{codegreen},
  keywordstyle=\color{magenta},
  numberstyle=\tiny\color{codegray},
  stringstyle=\color{codepurple},
  basicstyle=\ttfamily\footnotesize,
  breakatwhitespace=false,         
  breaklines=true,                 
  captionpos=b,                    
  keepspaces=true,                 
  numbers=left,                    
  numbersep=5pt,                  
  showspaces=false,                
  showstringspaces=false,
  showtabs=false,                  
  tabsize=2
}

\AtBeginDocument{%
  }

\acmYear{}
\acmDOI{}
\acmISBN{}
\acmConference{}{}{}

\begin{document}

\title[YATE: The Role of Test Repair in LLM-Based Unit Test Generation]{YATE: The Role of Test Repair in LLM-Based Unit Test Generation}

\author{Michael Konstantinou}
\email{michael.konstantinou@uni.lu}
\affiliation{
  \institution{University of Luxembourg}
  \country{Luxembourg}
}
\orcid{0009-0009-7726-178X}

\author{Renzo Degiovanni}
\email{renzo.degiovanni@list.lu}
\affiliation{
  \institution{Luxembourg Institute of
Science and Technology}
  \country{Luxembourg}
}
\orcid{0000-0003-1611-3969}

\author{Jie M. Zhang}
\email{jie.zhang@kcl.ac.uk}
\affiliation{%
  \institution{King’s College London}
  \country{United Kingdom}
}
\orcid{0000-0003-0481-7264}

\author{Mark Harman}
\email{mark.harman@ucl.ac.uk}
\affiliation{%
  \institution{University College London}
  \country{United Kingdom}
}
\orcid{0000-0002-5864-4488}

\author{Mike Papadakis}
\email{michail.papadakis@uni.lu}
\affiliation{%
  \institution{University of Luxembourg}
  \country{Luxembourg}
}
\orcid{0000-0003-1852-2547}

\renewcommand{\shortauthors}{Konstantinou et al.}

\begin{abstract}
Recent advances in automated test generation utilises language models to produce unit tests. While effective, language models tend to generate many incorrect tests with respect to both syntax and semantics. Although such incorrect tests can be easily detected and discarded, they constitute a ``missed opportunity'' -- if fixed, they are often valuable as they directly add testing value (they effectively target the underlying program logic to be tested) and indirectly form good seeds for generating additional tests. To this end, we propose a simple technique for repairing some of these incorrect tests through a combination of rule-based static analysis and re-prompting. We evaluate this simple approach, named YATE, on a set of 6 open-source projects and show that it can effectively produce tests that \textit{cover on average 32.06\% more lines and kill 21.77\% more mutants than a plain LLM-based method}. We also compare YATE with four other LLM-based methods, namely HITS, SYMPROMPT, TESTSPARK and COVERUP and show that it produces tests that cover substantially more code. YATE achieves \textit{22\% higher line coverage, 20\% higher branch coverage and kill 20\% more mutants} at a comparable cost (number of calls to LLMs).
\end{abstract}

\maketitle

\section{Introduction}

Recent advances in automated test generation have shown that language models can be effectively embedded into Software Engineering workflows to produce unit tests \cite{foster:mutation, mhetal:fse24-llm, chattester/10.1145/3660783, WangL0J24}. 
Although language models are prone to hallucination with detrimental affects on software engineering applications \cite{mhetal:LLM-survey}, Assured LLM Based Software Engineering (Assured LLMSE) approaches \cite{mhetal:intense24-keynote} provide assurances that the tests generated by the overall process are immediately deployable into production. 
This approach has already found effective industrial deployment at scale for both coverage based \cite{mhetal:fse24-llm} and mutation based \cite{foster:mutation} test generation.

Although suboptimal LLM performance (such as hallucination), can be mitigated using Assured LLMSE, the overall effectiveness and efficiency of test generation can be dramatically improved by tackling these issues using re-prompting (such as Chain of Thought and Self-Consistency) \cite{abs-2201-11903, 0002WSLCNCZ23}, retrieval augmentation (RAG) \cite{gao2023retrieval}, and hybrids of traditional Software Engineering, e.g., syntactic code analysis, and LLM-based test generation \cite{chattester/10.1145/3660783}.

LLMs may generate unit tests that are discarded by the assurance process due to either incorrect syntax or incorrect semantics.
When the syntax is incorrect, this can lead to a `near miss'; the test case generated might have proved valuable, but this potential value is lost simply because the test case syntax is not {\em quite} right (although it is {\em almost} right). Such near-miss syntactic errors include missing imports, and incorrect method invocations. We aim at repairing these near-miss tests and propose four techniques to repair relatively minor syntactic issues, using a combination of rule-based static analysis techniques and re-prompting with additional information to guide the model towards correct syntax.

However, even when the generated test is syntactically correct, there might remain semantic inadequacies to repair. 
In the case of unit tests, there are two primary semantic aspects of concern: 
the paths traversed by the test cases execution, which achieves a certain degree of coverage;  
and the assertions included as part of the test oracle \cite{ebetal:oracle} to determine whether the observed software behavior is the expected or not.

Many execution paths can be easily covered by arbitrary test method call invocations, making them a straightforward target for test generation engines. However, some specific execution paths can only be reached under particular conditions involving objects/methods other than the class under test, which need to be appropriately instantiated through relevant classes and method call invocations (aka test dependencies) \cite{FraserZ11}. To cover such cases, LLM-based test generation requires contextual information that resides outside the method/class under test. 

In the case of the test assertion(s), they may simply be semantically incorrect, leading to a test case that passes when it should fail or fails when it should pass.
However, determining whether a passing assertion should fail is known to be difficult \cite{mhpohss:harden}, known as the oracle problem \cite{ebetal:oracle}.
On the contrary, when applied in the context of `regression testing' \cite{alshahwan:software}, it is easier to fix an assertion that currently fails to make it pass, just by observing the actual program outputs.
Such an assertion fix is a relatively simple process that does not require further prompting to the LLM, and can produce a semantically correct test from a semantically incorrect test.

Even when the test is syntactically and semantically correct, i.e., a test that compiles and passes, the coverage and fault-detection achieved can still be weak. It is therefore natural to also consider re-prompting, retrieval augmentation and coverage feedback for improving the overall semantics of the test suite generated so that it maximally covers code and oracle aspects.

In this paper, we empirically study YATE, an LLM-based test generation technique that relies on four independent optimization components aiming at tackling the aforementioned syntactic and semantic shortcomings.  The four components studied are:

\begin{enumerate}
\item {\bf Initial test generation with chain-of-thought}: We instruct the LLM to identify the tests required to cover 100\% of the code (list the number of tests and their respective summary per class method targeted). In case the code under test depends on other classes, as well as on private or protected methods, YATE includes all this relevant information in the prompt in a follow-up step to ensure their consideration. Before actual test generation, the LLM is further challenged to reason on whether this set of tests will achieve 100\% branch coverage.

\item {\bf Compilation error Fixing}: When the generated test class does not compile, YATE identifies and fixes missing imports and constructors. If compilation errors remain, YATE recursively walks over the method call graph (up to a certain depth, similar to observation-based test generation \cite{mhetal:TestGen-obs}) to fetch classes and sub-types dependencies which are included in the prompt. Thus, YATE's compilation fixing uses re-prompting and retrieval augmented generation to fix incorrect method invocations.

\item {\bf Oracle Fixing}: When the tested assertion(s) currently fail(s), YATE uses a simple rule-based approach to fix assertions that can be made to pass by simple substitutions, such as expected/actual values and exception raising behaviours. 
Additionally, YATE's oracle fixing uses re-prompting with test failure logs to seek to regenerate the test assertions with a further call to be LLM. 

\item {\bf Coverage-Based Test Augmentation}: To improve test effectiveness, YATE re-prompts the LLM with information about uncovered branches.
\end{enumerate}

We empirically evaluate YATE's test effectiveness on 6 real-world Java projects, using 4 different LLMS, leading to a total of 393 test generation problems at both the method level and the class-level granularity. The subjects used in our evaluation are taken from the previous test generation studies of ChatUniTest \cite{chatunitest/10.1145/3663529.3663801}, and of HITS \cite{WangL0J24}, together with the Gitbug Java bug data set~\cite{SilvaSM24}. We control for data leakage, and select the benchmarks and LLMs accordingly their generation and training dates. We also include four LLM-based test generation tools (HITS~\cite{WangL0J24}, SymPrompt~\cite{10.1145/3643769}, TestSpark~\cite{sapozhnikov2024testspark} and Coverup~\cite{abs-2403-16218}) that we use as baselines in our experiments.

Our results show that YATE achieves an overall branch coverage of 51\%, which is substantially higher than the 30\%, 27\%, 17\% and 25\% coverage achieved by HITS, SymPrompt, TestSpark, and Coverup. We also investigate the individual contribution of each YATE's components by performing an ablation study. By excluding each one of the four optimization components listed above, YATE drops its branch coverage to 35\%, 35\%, 34\% and 40\%, respectively. These results show that even without one of the four components of YATE, it still outperforms all the state of the art methods studied. 

Our results also show that class level test generation offers a surprisingly favorable trade-off between test effectiveness and efficiency; 
coverage is only 5\% lower with  class level prompting (compared to method level prompting), yet consumes dramatically fewer resources (5x fewer LLM calls). 
Furthermore, the branch coverage achieved when prompting the LLM at the class level is 15.20\% higher than that achieved by the previous methods, while requiring 19,569 fewer LLM calls than them. 

In summary, the key scientific conclusions of this paper are:

\begin{itemize}
\item {\bf Fixing 'near-miss' broken tests pay-off}:   
YATE implements four test-repair components that, when combined, achieve a branch coverage of 51\%, on average, surpassing state-of-the-art tools.
\item {\bf Class-level prompting is more cost-effective}: our results reveal that prompting at the class level is surprisingly effective {\em and} efficient. Compared to the state-of-the-art, class level prompting achieved up to 73.22\% effectiveness improvement combined with 82\% efficiency improvement. 
\item {\bf Consistent observations through different LLMs}: Although certainly some models perform better than others, we can observe that YATE's performance is consistently improved when the test-repair components are activated. 
This suggests that each of the individual test-repair components by itself is effective, and can help to improve other tools effectiveness as well.
\item {\bf Compilation-relevant feedback is imperative}: We also find that providing LLMs with compilation-relevant feedback, such as class dependencies, is more important, wrt test effectiveness, than coverage feedback. 
\end{itemize}

\section{Related Work}

ChatTester is one of the first studies that used LLMs to generate unit tests \cite{chattester/10.1145/3660783}. ChatTester simply asks LLMs to generate tests for a given class and then iteratively improves them by providing feedback (the compiler error message) every time an error is thrown, i.e., when the test is not compiling. This scheme improves the test suites by generating 34.30\% more compiling tests and 18.7\% more passing tests. This feedback-driven repair loop was later adopted by many LLM-based unit test generators, implementing variations of feedback-driven repair~\cite{pan2024aster}~\cite{abs-2503-14000}~\cite{11042526}~\cite{abs-2506-02943}~\cite{SchaferNET24}~\cite{AlshahwanCFGHHM24}~\cite{NanG0025}~\cite{abs-2406-15743} using compiler error messages to make LLMs fix the tests they generate. YATE goes a step further in the compilation fixing by providing relevant context, test dependencies, and directly fixing asserted variable values, imports, constructors, and incorrect method invocations.

TestSpark~\cite{sapozhnikov2024testspark} aimed to improve the prompts that were used by enriching them with some context, i.e., included the class under test together with the relevant parent classes. CoverUp~\cite{abs-2403-16218} introduced the idea of providing coverage feedback, i.e., uncovered code, to LLMs so that they can effectively generate tests for uncovered code. Similarly to the aforementioned approaches, CoverUp identifies the code segments (function or class), that require testing, uses the LLM to generate tests and provides the error messages and the uncovered code as feedback (re-prompts the LLMs) for a number of iterations. 
ChatUniTest~\cite{chatunitest/10.1145/3663529.3663801} includes a test repair mechanism for simple syntactic errors such as imports and missing semicolons, which are taken from the compilation error message. Although conceptually similar to YATE, the test fixing phase of ChatUniTest remains relatively weak because it does not aim at providing relevant content, and does not account for test dependencies such as incorrect method invocations and calls. 

SymPrompt~\cite{10.1145/3643769} prompts LLMs with information drawn from traditional path-based code analysis. SymPrompt statically analyzes a method to identify all feasible execution paths, and then prompts a large language model (LLM) to generate a corresponding test case for each path. PANTA~\cite{abs-2503-13580} identifies uncovered code segments and iteratively prompts the LLMs with candidate execution paths as the target. Although it's repairing mechanism doesn't differ much from other approaches, it does keep a list of failed tests that are provided as feedback to the next test generation.

HITS~\cite{WangL0J24} is a tool that goes deeper into a method before generating tests through slicing. Using the "Chain of Thought" prompting strategy, it instructs the language model to decompose a given method into smaller, logically coherent code blocks known as slices. The tool then prompts the model to generate a corresponding test class for each slice.

TELPA~\cite{abs-2404-04966} performs a method invocation analysis to identify method call sequences. The methods involved in these sequences are then used as context in the prompt that asks for test generation. In parallel to YATE, these approaches aim at reducing the context that is given to the LLM (as a 0-shot generation) and not to iteratively guide them towards fixing and improving the tests they generate as done by YATE. 

There are also some approaches aiming to generate unit tests to \textit{kill} \textit{mutants} \cite{DBLP:conf/icst/StraubingerKL025}~\cite{abs-2506-02954}. Notably, Mutap~\cite{DBLP:journals/infsof/DakhelNMKD24}, ACH~\cite{abs-2501-12862}, PRIMG~\cite{DBLP:journals/corr/abs-2505-05584} are some of the approaches that have been used to generate unit tests to improve the mutation score of generated tests. These approaches differ from the above ones by providing mutation feedback (live mutants) instead of code coverage. 

Perhaps the closest approach to ours is TestART~\cite{abs-2408-03095}, that uses template-based and re-prompting to perform test compilation fixing. It relies on five templates to fix common compilation issues. However, while YATE aims at both test and oracle fixing and re-prompting with test dependencies, TestART focuses on incorrect oracles only, and its fix templates are too specific to Junit4.

Overall, each of the above approaches focuses on different aspects involved in the LLM unit test generation process. ChatTester and ChatUniTest focus on making tests to compile by iteratively repairing the generated tests, but their fixing strategies are superficial, as they are solely based on compiler error messages living out vital contextual information that the LLMs lack, such as relevant constructors, classes, method declarations, and method call signatures. Studies like Hits or SymPrompt focus on effective zero-shot test generation by tyring to make their prompts coherent and relevant by including information such as slices and target execution paths thereby leaving out the test repair and the coverage guidance/feedback aspects of the test generation process. Approaches like TestART aim to fix the generated assertions, and CoverUp to cover more code through coverage feedback. YATE shows that although all phases (relevant information in the prompt, test and assertion fixing, coverage guidance) are important, a general repairing mechanism that allows one to fix the generated tests is vital and largely overlooked by previous studies. YATE outperforms all previous approaches by using syntactic checks and incrementally providing the LLM the relevant content needed to fix its tests through dependence analysis.

\section{The YATE approach}

\subsection{Overview}

The main strategy followed in LLM-based test generation is to provide the code of the class (or a method) under test based on which the LLM will generate the tests. When the initial attempt fails, a sequence of feedback-based refinement prompts (with error messages or uncovered code) is typically used to guide the model toward a correct solution.

Although to some extent successful, the key problem with the above scheme is that classes (and their respective tests) have dependencies outside the class under test. Therefore, it is unlikely that an LLM will manage to generate valid tests that effectively utilize those dependencies without any prior knowledge of them. 

Consider the example below (listing \ref{lst:pay-constructor}) where the class under test (\texttt{Pay.java}) instantiates object \texttt{RequestHandler} in the constructor. In this example, the constructor of \texttt{Pay.java} requires two String variables, one boolean and two objects, namely \texttt{SignatureGenerator} and \texttt{ProxyAuth}. Both objects belong to the project under test, but are stored under different namespaces, whereas \texttt{SignatureGenerator} is an interface. If the LLM has not been provided with the definition of \texttt{ProxyAuth} and a class that implements \texttt{SignatureGenerator}, it is unlikely to figure out how to correctly instantiate an object of such a type. Therefore, YATE analyzes the repository of the code under test in order to provide to the model information about the constructors of \texttt{ProxyAuth} and the constructors of the classes that implement the interface \texttt{SignatureGenerator}.

\begin{lstlisting}[language=Java, caption=Constructor of class Pay.java in repository binance-connector-client, label=lst:pay-constructor]
package com.binance.connector.client.impl.spot;

/**
 * One of the two constructors of class Pay
 * SignatureGenerator is an interface
**/
public Pay(String baseUrl,
       String apiKey,
       SignatureGenerator signatureGenerator, 
       boolean showLimitUsage,
       ProxyAuth proxy) {
    this.baseUrl = baseUrl;
    this.requestHandler = new RequestHandler(apiKey, signatureGenerator, proxy);
    this.showLimitUsage = showLimitUsage;
}
\end{lstlisting}

\subsection{Workflow}

\begin{figure*}
    \centering
    \vspace{-1.0em}
    \includegraphics[width=0.95\linewidth]{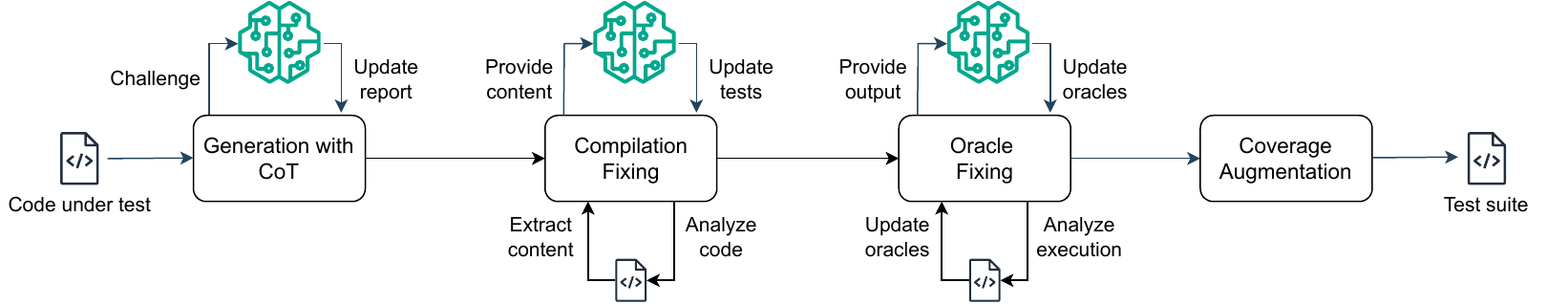}
    \vspace{-0.5em}
    \caption{YATE workflow, based on the four components}
    \label{fig:workflow}
\end{figure*}
Overall, the YATE's approach is to mimic the developer's workflow and its workflow consists of the following four key components. 

\begin{itemize}
    \item Generation of initial test code through chain-of-thought.
    \item Fixing compilation errors. This step involves the automatic debugging of compile errors, the analysis of the project under test, the fetching of relevant content, the finding of possible fixes to compile errors and the iterative and incremental re-prompting with relevant content to the encountered compiler errors.
    \item Fixing of incorrect oracles based on the actual program behavior, i.e., the actual runtime program state such as the actual program variable values of the program. 
    \item Coverage feedback. Looking for uncovered branches and based on the generated test class, the LLM is instructed to generate tests for the missing branches.
\end{itemize}

It should be noted, that YATE is extensible, meaning that one can replace each component with a different implementation to enhance the unit test generation process. 

\subsection{Initial test generation phase}

Before generating any test cases, YATE prompts the language model to analyze the code under test (class or method) with the objective of achieving 100\% code coverage and to produce a corresponding report that includes a list with the description of the test cases required per method, and the total number of required tests. This initial step functions as a summarisation of the required test cases and guides the model to reason more effectively about the required tests, as suggested by the Chain of Thought~\cite{abs-2201-11903} prompting strategies. If the class under test includes private or protected methods, YATE issues an additional prompt to inform the model of these methods, subsequently updating the report to reflect this information. Before instructing the model to generate tests, an additional prompt is used, asking the LLM to verify whether the current report will achieve branch coverage 100\%, to further challenge the LLM. This additional prompt drives the model to reason more effectively about the required tests as suggested by several prompting guidelines. Once the report phase is completed, YATE proceeds to instruct the model to generate the corresponding test code.

\subsection{Fixing compilation errors}

In this phase YATE executes the generated tests and examines the error log. If the tests do not compile, YATE attempts to incrementally fix the generated tests by providing fixes for some simple compilation error cases and by providing relevant content to guide the LLMs to fix the errors.  
YATE's compilation fixing is based on a breadth-first approach. It begins with issues that affect the whole file, then incrementally fixes issues that affect all test cases, and focuses on the content of the broken tests. 

\subsubsection{Missing imports}

Typically, the class under test and its respective unit tests use libraries or other classes from the repository. These need to be imported into the test class in order to compile. Related work primarily operates at the method granularity and thus provides the LLMs only the implementation of the method under test. This causes many issues with missing import statements or with irrelevant or non-existent import statements, i.e., the import statements are a product of hallucinations. 

To deal with these cases, YATE identifies the test classes without the proper imports and statically analyses the repository of the class under test to find the import statements that are relevant to the classes used by the tests and directly imports them. To fix imports that are hallucinations, YATE analyzes the import statements by checking whether the reference exists, e.g., checking if the imported class exists in the project repository or relevant classpath space.

\subsubsection{Retrieving the implementation of classes other than the class under test}

The generated test class often constructs objects the code of which are in a different file/class. YATE searches for the required classes and collects all constructor signatures. In case a required object is an interface or an abstract class, YATE retrieves their code classes. 
YATE then provides this information to the LLM and instructs it to fix its tests based on a new content.

\subsubsection{Wrong method invocations}

LLMs often generate test code with method calls that do not exist, or methods with wrong input parameters. This may well extend to incorrectly mocking method calls, or their returned value.  If a method call regards a method does not exist, YATE points this out to the LLM. If the method exists but is incorrectly invoked, the LLM is provided with the method signatures of those methods. Note that this step applies only to the wrong test invocations appearing in the generated test classes.  

\subsubsection{Method calls outside the class under test}

If errors persist, YATE uses the project's method call graph and searches for method invocations in the scope of the method under test that belong to other files/classes. The bodies of these methods (methods invoked outside the class under test) along with their signatures are then included in a follow-up prompt. Additionally, the prompt explicitly informs the language model about the relationships between the methods under test and the external methods they invoke. Note that this step retrieves code from the methods called by the methods of the class under test, not the generated test class(es). This step provides the LLM information on the functionality of methods under test by flattening the related code that is shown to the LLMs.

\subsubsection{Error log as feedback between the previous steps}

After every one of the previous steps, if LLMs have made any changes to their tests but failed to compile them, YATE provides the complete error log. During our experiments, we found that unambiguous constructor calls are a recurrent issue, which can be solved by providing stricter-typed variables, e.g., providing the object instead of null. Therefore, in these cases, YATE prompts the LLMs to fix the generated tests by providing the related object (instead of null). Similarly, other recurrent compilation errors can be fixed. 

\subsection{Oracle fixing}

In the oracle fixing phase, YATE makes the following three  steps:

\subsubsection{Correcting incorrect values}

Initially, YATE attempts to fix the oracles (test assertions) based on the execution outcome (runtime program state). This process is in line with the process followed by both LLM-based, e.g., TestART ~\cite{abs-2408-03095}, and traditional, e.g., Evosuite \cite{6004309} test generation approaches. 

YATE attempts to fix primitive values, like numbers, text, boolean adn etc., simply by recording the variable values involved in the generated assertions. The rules used to fix the oracles are: 

\begin{itemize}
    \item Explicit values such as strings or numbers, are immediately replaced with the test execution's outcome.
    \item Non-explicit values such as assertTrue, or assertNotNull, are reversed to reflect the execution outcome.
\end{itemize}

\subsubsection{Generating or inverting exception oracles}

In case the oracle is failing due to the fact that the test is throwing an exception, then YATE will replace the last oracle with an exception one that verifies the thrown exception. It is possible that sometimes the thrown exception is occurring due to the wrong implementation of the unit test. To avoid such issues, YATE has a configurable list with exception oracles to avoid. More specifically, by default, YATE will not generate an exception oracle if the exception type is related to incorrect mock usage. Similarly, YATE will check for exception oracles that did not throw exception in their executions, and invert the exception oracle. For instance, in Java's Junit5 testing framework, an assertThrows oracle will become assertDoesNotThrow.

This step is also rule-based, but if the current (exception) oracle is complicated enough to parse, YATE will provide the oracle line to the LLM and instruct it to make the desired modifications.

\subsubsection{Fixing complex values using an LLM}

If none of the above steps were enough to fix the oracles, YATE will provide to the LLM the error log and instruct it to fix the oracles based on the test execution. This process can be repeated for a configurable N number of iterations or until the oracles are fixed. By default, the maximum number of iterations is 3. 

If any of the aforementioned changes stops the test case from compiling, then the last change that caused this issue is reverted. 

\subsection{Coverage-based test augmentation}

Once the test generation process is complete, YATE executes the test cases and measures their coverage. In case YATE finds any uncovered branches, then it askes the LLMs to generate a new test class to generate unit tests that cover the missed branches. 

YATE does this by starting a new process in which it provides to the LLM the class implementation, the generated test class implementation, and the missed branches. The repairing mechanism of the two components "Compilation fixing" and "Oracle fixing" is then incorporated as presented above. The process stops at the end of this (second) generation cycle.

\section{Experimental Design}

\subsection{Research Questions}

We start our analysis by investigating the ability of YATE to generate tests that compile. We contrast the compilation and pass rates, as well as the actual number of the tests generated by YATE with those generated by a simple (plain) call to an LLM, to show the improvements introduced by YATE. Therefore, our first research question can be formulated as follows.

\textit{\textbf{RQ1. Syntactically valid Tests:} How does YATE compare to a plain LLM-based test generation in terms of the number of syntactically valid tests generated and the test compilation and pass rates?}

To answer this question, we select a set of projects, entire repositories, where we apply the plain LLM-based and YATE test generation methods and contrast the number of tests that compile and pass. Since the plain method forms the first step of YATE, the difference in these numbers is attributed to the YATE compilation and oracle fixing steps. We distinguish between class and method granularity levels, by providing the LLM the entire class under test or only the method definition. The reasoning behind this distinction is to check whether the class under test includes additional content that could be useful to LLMs. 

After verifying compilation and pass rates on the selected projects in RQ1, we check whether the generated test suites have significant differences in terms of test effectiveness. The effectiveness is measured based on typically used test adequacy metrics \cite{ZhuHM97}, such as line and branch coverage and mutation testing \cite{PapadakisK00TH19}. Hence, our second research question can be formulated as follows.

\textit{\textbf{RQ2. YATE Test Effectiveness:} How does YATE compare to a plain LLM-based test generation in terms of test effectiveness?}

Similarly to RQ1 we measure test effectiveness and contrast it to that of a plain LLM-based method on the selected projects by distinguishing between class and method granularity levels. 

Answering the first two questions provides evidence related to the overall effectiveness of YATE but does not ensure that all its steps contribute effectively to its performance. We therefore investigate through an ablation study whether all 4 optimizations we consider are useful, i.e., we check whether they help the YATE test generation method produce stronger tests than those it would produce if any of those optimizations were absent. Thus, we ask:

\textit{\textbf{RQ3. Ablation study:} Do all considered optimizations (chain-of-thought prompt, Compilation Fixing, Oracle Fixing, and Coverage-based Test Augmentation) contribute to YATE's performance?}

Answering this question helps us understand the contribution of each of the 4 optimizations in YATE's performance. However, since all of them are strongly coupled with the abilities/performance of the considered LLMs to perform the requested tasks, we further investigate the sensitivity of YATE to different LLMs. Thus, we ask:

\textit{\textbf{RQ4. Sensitivity to LLMs used:} How effective is the YATE test generation method when using different LLMs?}

To answer this question we use GPT-4o-mini~\cite{openai2024gpt4omini}, GPT 4.1,~\cite{openai2025gpt41} Deepseek V3~\cite{abs-2412-19437} and Codestral~\cite{mistral2024codestral}, which are the most recent LLMs to date. After checking YATE's performance, we compare YATE against existing methods. Thus, we contrast YATE's performance with that of HITS \cite{WangL0J24}, SYMPROMPT ~\cite{10.1145/3643769}, TESTSPARK \cite{sapozhnikov2024testspark}, and COVERUP~\cite{abs-2403-16218} and ask:

\textit{\textbf{RQ5. Comparison with SOTA:} How does YATE compare to SOTA approaches in terms of test effectiveness and execution performance?}

We answer this question by measuring both the cost and the effectiveness of the methods studied. Effectiveness is measured based on coverage and mutation testing, while cost is measured based on the number of LLM calls made by YATE, which constitutes the key cost contributing factor of LLM-based test generation methods~\cite{chattester/10.1145/3660783}.

\subsection{Projects used}

To evaluate our approach, we selected a set of projects also used by previous studies for evaluating ChatUniTest~\cite{chatunitest/10.1145/3663529.3663801} and HITS~\cite{WangL0J24}, as well as the Gitbug java bug dataset~\cite{SilvaSM24}.

We also selected some projects previously seen by LLMs and some others on which LLMs have not been trained to check for performance differences. To distinguish between seen and unseen data, we follow a strict cutoff date: any project created after October 1, 2023, is considered unseen, under the assumption that such data were not included in the pre-training corpus of GPT-4o-mini, which was the primary model used across all research questions due to its superior performance. All other projects are labeled as seen. However, it should be noted that for the last research question, some LLMs may have seen different sets of projects.

Notice that we run our evaluation on all classes from the selected projects to avoid favoring any method due to data selection. This is in contrast to what was done in previous work that was evaluated on self-selected sets of classes or methods.

\begin{table}[t!]
\centering
\caption{Projects used}
\vspace{-1.0em}
\resizebox{\columnwidth}{!}{
\begin{tabular}{ccccccl}
\toprule
\textbf{Project} & \textbf{Abbr} & \textbf{JDK} & \textbf{\# CUT} & \textbf{\# MUT} & \textbf{Previously used} & \textbf{Seen} \\
\midrule
Binance / Binance connector 2.0.0 & BC  & 8  & 46  & 481  & ChatUniTest & Yes \\
bhlangonijr / Chesslib    & CH  & 11 & 43  & 538  & Gitbug Java & Yes \\
AuthMe / ConfigMe       & CM  & 8  & 75  & 541  & Gitbug Java & Yes \\
AWS / Event-ruler             & ER  & 8  & 53  & 656  & HITS        & No  \\
Flmelody / Windward                & W   & 8  & 78  & 400  & HITS        & No  \\
Apple / Batch-processing-gateway& BPG & 17 & 98  & 1041 & HITS        & No  \\
\midrule
\textbf{TOTAL} &       &     & \textbf{393} & \textbf{3657} &  & \\
\bottomrule
    \vspace{-1.0em}
\end{tabular}
}
\end{table}

\subsection{Baselines}

To study the effectiveness of YATE, we evaluated it against a simple \textit{LLM-based plain baseline}. This baseline consists of one prompt that aims to generate unit tests, followed by a maximum of five iterations with the LLM using the error logs in the prompts so that the LLM can attempt to fix the tests that do not compile. It is also equipped with a rule-based fix of the import statements in order to eliminate any advantage attributed to such trivial cases.

We also compare YATE with the following state-of-the-art LLM-based test generation methods. Each baseline and YATE were executed using GPT-4o-mini~\cite{openai2024gpt4omini} to ensure a consistent and fair comparison.

\textit{HITS}~\cite{WangL0J24} uses LLMs to cut methods into smaller code blocks before testing a class. Afterwards, HITS uses the LLM to generate unit tests for each sliced code block. 

\textit{SymPrompt}~\cite{10.1145/3643769} mimics symbolic execution. Identifies all possible execution paths and instructs the LLM to generate unit tests for each execution path. 

\textit{TestSpark}~\cite{sapozhnikov2024testspark} contains an LLM-based implementation and an Evosuite-based implementation. Since we conduct an empirical study on LLM-based unit test generation, we used the TestSpark's LLM-based approach. TestSpark instructs the LLM to generate unit tests to achieve 100\% branch coverage and is the only approach that provides also the parent class(es) of the class under test. 

\textit{Coverup}~\cite{abs-2403-16218} attempts to improve the initial LLM-based test generation, by identifying uncovered lines and generating additional tests to cover them. 

Unfortunately, we could not find any working implementations for the ASTER~\cite{pan2024aster}, PANTA~\cite{abs-2503-13580} and TELPA~\cite{abs-2404-04966} approaches, and thus we did not consider them. Nevertheless, as we discuss in the related work, these approaches are somehow orthogonal to ours as they do not aim at fixing tests using contextual information. TestART~\cite{abs-2408-03095} could not be used as it does not supports Junit5 that is used by our projects.

\subsection{Metrics}

To evaluate \textit{test effectiveness}, we used standard test adequacy metrics (aka Code coverage metrics) \cite{ZhuHM97}, such as line and branch coverage measured using the Jacoco\footnote{https://www.jacoco.org/jacoco/} framework. We also calculated the Mutation Score achieved by the generated test suites using the default PiTest configuration~\cite{ColesLHPV16}. 

To measure efficiency and \textit{approximate the cost }of the studied test generations methods, we count the number of requests that each approach makes to the LLMs employed. We consider this number as a good cost measurement since the LLM queries form the most expensive (computationally) part of the studied approaches~\cite{chattester/10.1145/3660783}. Moreover, the LLM queries also form an actual cost factor, specially for the commercial ones. 

It is noted that the model queries may fail due to unexpected errors, such as server exception, API errors etc. To account for these factors, in our experiments we repeated those queries for a number of times but only counted them as one request, i.e., without counting repetitions due to such errors. 

\textbf{Excluding invalid tests:} Non-compiling or non-passing tests are filtered out in order to be able to calculate the coverage of each approach. All the approaches studied attempt to filter out invalid tests. However, in some error cases we may not be able to filter all invalid tests, such as the cases where we cannot parse the generated test code. Therefore, invalid test code was filtered as follows:
\begin{itemize}
    \item All non-compiling tests or methods are removed. This applies also to assistant methods generated in the test class and not only non-compiling tests.
    \item If a test class does not compile as a whole (regardless of the test cases), then the whole file is removed.
\end{itemize}

\textbf{Comparisons:} To compare whether the differences between the results are statistically significant, we apply the Mann-Whitney U test~\cite{mann1947test}, a non-parametric test for comparing two independent samples. In our analysis, we consider each class under test as a data point and compare the distributions of line and branch coverage. We report the corresponding p-value, considering the results statistically significant when p < 0.05. To evaluate the magnitude of difference between observed groups, we also calculate the Vargha and Delaney (VDA) effect size~\cite{VarghaDelaney2000}.

Alongside statistical significance, we report two additional metrics to compare coverage results. We use \textbf{coverage difference} which is the absolute percentage point increase in coverage and the coverage improvement. 

\section{Results}

\subsection{RQ1: Compilation and Pass rates}

Tables \ref{tab:compilation-rates} and \ref{tab:passing-rates} report the compilation and passing rates of tests generated by LLM-Plain and YATE, for both class and method granularity levels. In both cases, YATE shows a substantial improvement in compilation and passing rates compared to LLM-Plain. At the class-level, YATE achieved a compilation rate of 94.60\%, which is 5.96\% higher than the compilation rate of LLM-Plain. For method-level tests, the YATE compilation rate was 90.02\%, surpassing the LLM-Plain by 5.68\%. 

When looking at the passing rate of YATE's tests, we observe that the same conclusions can be drawn, and the difference is much higher. Although LLM-Plain's passing rates hovered just above 50\% for both configurations, YATE achieved 90.18\% for class-level and 82.96\% for method-level testing in total. 

\begin{table}[t]
\centering
\caption{Compilation Rates of YATE and LLM-Plain}
    \vspace{-1.0em}
\resizebox{\columnwidth}{!}{
\begin{tabular}{lcccc}
\toprule
\textbf{Project} & PLAIN-C & PLAIN-M & YATE-C & YATE-M \\
\midrule
BPG & 87.19\% (640/734) & 87.77\% (2714/3092) & \textbf{93.68\% (1097/1171)} & \textbf{92.36\% (2551/2762)} \\
ER  & 77.66\% (431/555) & 69.91\% (1705/2439) & \textbf{95.09\% (871/916)} & \textbf{73.91\% (1904/2576)} \\
BC  & 98.83\% (509/515) & 93.91\% (1572/1674) & \textbf{99.05\% (832/840)} & \textbf{99.43\% (1754/1764)} \\
W   & 87.13\% (474/544) & 83.01\% (850/1024) & \textbf{98.31\% (641/652)} & \textbf{96.00\% (1079/1124)} \\
CM  & 86.89\% (444/511) & 80.54\% (1291/1603) & \textbf{86.90\% (776/893)} & \textbf{90.75\% (1462/1611)} \\
CH  & \textbf{98.39\% (367/373)} & 91.23\% (2122/2326) & 97.48\% (464/476) & \textbf{96.23\% (1633/1697)} \\
\midrule
Total & 88.64\% (2865/3232) & 84.34\% (10254/12158) & \textbf{94.60\% (4681/4948)} & \textbf{90.02\% (10383/11534)} \\
Average & 89.35\% & 84.39\% & \textbf{95.08\%} & \textbf{91.45\%} \\
\bottomrule
\end{tabular}
}
\label{tab:compilation-rates}
\end{table}

\begin{table}[t]
\centering
\caption{Passing Rates of YATE and LLM-Plain}
    \vspace{-1.0em}
\resizebox{\columnwidth}{!}{
\begin{tabular}{lcccc}
\toprule
\textbf{Project} & PLAIN-C & PLAIN-M & YATE-C & YATE-M \\
\midrule
BPG & 71.25\% (523/734) & 76.13\% (2354/3092) & \textbf{87.45\% (1024/1171)} & \textbf{88.38\% (2441/2762)} \\
ER  & 42.70\% (237/555) & 37.39\% (912/2439) & \textbf{92.36\% (846/916)} & \textbf{65.84\% (1696/2576)} \\
BC  & 14.17\% (73/515) & 19.83\% (332/1674) & \textbf{98.21\% (825/840)} & \textbf{79.88\% (1409/1764)} \\
W   & 61.76\% (336/544) & 47.95\% (491/1024) & \textbf{92.48\% (603/652)} & \textbf{93.68\% (1053/1124)} \\
CM  & 42.07\% (215/511) & 47.72\% (765/1603) & \textbf{83.54\% (746/893)} & \textbf{84.48\% (1361/1611)} \\
CH  & 66.76\% (249/373) & 70.42\% (1638/2326) & \textbf{87.82\% (418/476)} & \textbf{94.81\% (1609/1697)} \\
\midrule
Total & 50.53\% (1633/3232) & 53.40\% (6492/12158) & \textbf{90.18\% (4462/4948)} & \textbf{82.96\% (9569/11534)} \\
Average & 49.79\% & 49.91\% & \textbf{90.31\%} & \textbf{84.51\%} \\
\bottomrule
\end{tabular}
}
\label{tab:passing-rates}
\end{table}

In terms of actual numbers, LLM-Plain (class-level) generated 3,232 tests, of which 2,865 compiled and 1,633 passed. In contrast, YATE (class-level) generated 4,948 tests, with 4,681 compiling and 4,462 passing. This means that YATE generated 1,716 more syntactically valid tests and 2,829 more passing tests than LLM-Plain. For method-level testing, both approaches generated a larger number of tests. LLM-Plain generated 12,158 tests, with 10,254 compiling and 6,492 passing. YATE generated fewer tests in total (11,534) but achieved better compilation and passing rates, with 10,383 tests compiling and 9,569 passing. 

To summarize these results demonstrate that YATE generates a larger number of syntactically valid and passing tests, making YATE's test outputs more reliable on both granularity levels studied. 

\begin{tcolorbox}[colback=gray!10, colframe=gray!50,boxrule=0.4pt,arc=1pt,left=6pt,right=6pt,top=4pt,bottom=4pt,fontupper=\small,before skip=8pt,after skip=8pt,enhanced]
\textbf{Finding 1:} YATE significantly improves the compilation and passing rates of LLM-Plain. It generates 5.96\% more compiling tests and 39.64\% more passing tests, ultimately yielding a larger number of passing test cases in both configurations.
\end{tcolorbox}

\subsection{RQ2: Test Effectiveness}

\begin{table*}[t!]
\centering
\vspace{-0.4em}
\caption{Comparison of LLM Plain (PL-C, PL-M) and YATE (Y-C, Y-M) in 6 projects (BPG, ER, BC, W, CM, CH).}
\vspace{-1.1em}
\resizebox{\textwidth}{!}{
\begin{tabular}{lcccc|c|cccc|c|cccc|c}
\toprule
\textbf{Project} & 
\multicolumn{5}{c}{\textbf{Line Coverage}} &
\multicolumn{5}{c}{\textbf{Branch Coverage}} &
\multicolumn{5}{c}{\textbf{Mutation Score}} \\
\cmidrule(r){2-6} \cmidrule(r){7-11} \cmidrule(r){12-16}
& PL-C & PL-M & \textbf{Y-C} & \textbf{Y-M} & \textbf{Y-Comb} 
& PL-C & PL-M & \textbf{Y-C} & \textbf{Y-M} & \textbf{Y-Comb} 
& PL-C & PL-M & \textbf{Y-C} & \textbf{Y-M} & \textbf{Y-Comb} \\
\midrule
BPG & 36.47\% & 45.13\% & \textbf{44.71\%} & \textbf{50.88\%} & 54.34\% 
    & 22.88\% & 32.60\% & \textbf{32.84\%} & \textbf{41.30\%} & 46.00\% 
    & 29.33\% & 41.21\% & \textbf{34.25\%} & \textbf{45.71\%} & 45.71\% \\
ER & 35.58\% & 45.55\% & \textbf{58.00\%} & \textbf{58.98\%} & 67.44\%
    & 23.36\% & 32.10\% & \textbf{44.20\%} & \textbf{46.18\%} & 53.87\%
    & 23.03\% & 31.53\% & \textbf{43.17\%} & \textbf{42.07\%} & 50.30\% \\
BC  & 15.21\% & 49.91\% & \textbf{82.10\%} & \textbf{86.85\%} & 90.62\%
    & 36.20\% & 49.77\% & \textbf{80.82\%} & \textbf{84.47\%} & 86.30\%
    & 7.85\%  & 15.93\% & \textbf{51.76\%} & \textbf{41.10\%} & 44.73\% \\
W   & 37.14\% & 36.96\% & \textbf{63.54\%} & \textbf{66.02\%} & 71.49\%
    & 24.90\% & 27.49\% & \textbf{55.18\%} & \textbf{61.55\%} & 67.73\%
    & 30.48\% & 32.81\% & \textbf{50.86\%} & \textbf{58.94\%} & 63.61\% \\
CM  & 41.62\% & 65.39\% & \textbf{75.58\%} & \textbf{82.40\%} & 88.77\%
    & 25.56\% & 48.89\% & \textbf{63.52\%} & \textbf{81.48\%} & 88.33\%
    & 28.50\% & 53.24\% & \textbf{52.72\%} & \textbf{72.67\%} & 76.94\% \\
CH  & 31.24\% & 61.86\% & \textbf{65.69\%} & \textbf{69.82\%} & 73.99\%
    & 8.57\%  & 39.10\% & \textbf{37.02\%} & \textbf{47.12\%} & 51.28\%
    & 14.11\% & 30.00\% & \textbf{31.14\%} & \textbf{46.13\%} & 48.35\% \\
\midrule
\textbf{Total} & 33.53\% & 49.95\% & \textbf{60.71\%} & \textbf{64.82\%} & 70.13\%
              & 21.05\% & 35.33\% & \textbf{44.27\%} & \textbf{51.06\%} & 57.01\%
              & 22.08\% & 33.82\% & \textbf{41.39\%} & \textbf{47.99\%} & 52.04\% \\
\textbf{Average} & 32.88\% & 50.80\% & \textbf{64.94\%} & \textbf{69.33\%} & 74.44\%
                & 23.58\% & 38.33\% & \textbf{52.26\%} & \textbf{60.35\%} & 65.59\%
                & 22.22\% & 34.12\% & \textbf{43.98\%} & \textbf{51.10\%} & 54.94\% \\
\bottomrule
\vspace{-1.5em}
\end{tabular}
}
\label{tab:llmplain_vs_yate}
\end{table*}

Table \ref{tab:llmplain_vs_yate} presents the coverage achieved by the LLM Plain and YATE methods in all the studied projects. The table includes both class-level and method-level configurations, and additionally reports the overall coverage when class-level and method-level tests generated by YATE are combined into a single test suite.

We observe that models perform better when instructed to generate tests per method rather than generating tests per class. This happens because the models seem to generate more tests per method, allowing to test a method deeper. This is also shown in tables \ref{tab:compilation-rates} and \ref{tab:passing-rates} where both approaches generated more tests at the method level. Table \ref{tab:comparison-plain} records the results of the statistical analysis of the effectiveness values obtained per class involved.

We observe that the differences between class and method levels of the plain and YATE  approaches are statistically significant in all cases except from the case YATE class vs YATE method. The Vargha and Delaney effect size show that YATE (class) achieved higher line coverage in approximately 61.6\% of the classes under test, and higher branch coverage in about 69.6\% of the cases. A similar trend is observed when comparing YATE (method) to LLM-Plain (method), indicating that YATE (method) also achieves consistently better coverage.

\begin{tcolorbox}[colback=gray!10, colframe=gray!50,boxrule=0.4pt,arc=1pt,left=6pt,right=6pt,top=4pt,bottom=4pt,fontupper=\small,before skip=8pt,after skip=8pt,enhanced]
\textbf{Finding 2:} LLM's generate more syntactically valid tests and cover more code when they are instructed to focus on each method separately rather than generate tests for the whole class.
\end{tcolorbox}

\begin{figure}[b]
    \centering
    \vspace{-1.0em}
    \includegraphics[width=0.95\linewidth]{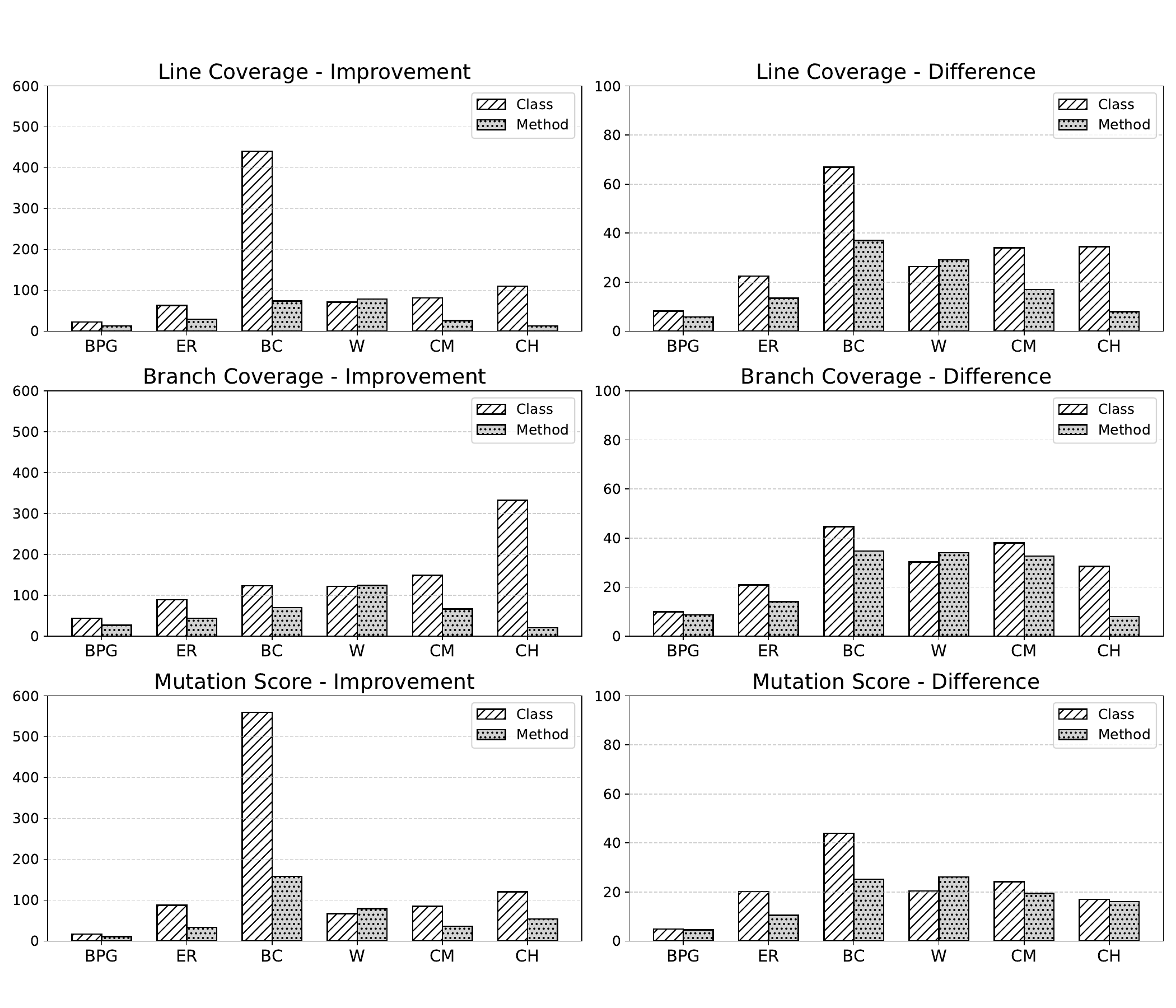}
    \vspace{-1.0em}
    \caption{Coverage improvement of YATE over LLM-Plain for the 6 projects (BPG, ER, BC, W, CM, CH) used and the three effectiveness metrics (Line, Branch, and Mutation Score)}
    \label{fig:impr_diff}
\end{figure}

Figure \ref{fig:impr_diff} shows the improvement in coverage when using YATE and the average difference in coverage compared to LLM-Plain. When used at class granularity level YATE improved line coverage by 131.45\%, branch coverage by 143.02\%, and mutation score by 156.07\%. On average, this translates to 32.06\% higher line coverage, 28.69\% higher branch coverage and 21.77\% higher mutation score compared to LLM-Plain. Although the improvements are smaller under the method-level configuration, YATE still outperforms LLM-Plain across all metrics. On average, YATE achieved 18.36\% more line coverage, 22. 03\% more branch coverage, and 16. 98\% more mutants killed. 

\begin{tcolorbox}[colback=gray!10, colframe=gray!50,boxrule=0.4pt,arc=1pt,left=6pt,right=6pt,top=4pt,bottom=4pt,fontupper=\small,before skip=8pt,after skip=8pt,enhanced]
\textbf{Finding 3:} YATE consistently achieves significantly higher code coverage than LLM-Plain, under both class-level and method-level test generation settings.
\end{tcolorbox}

Interestingly, the combination of YATE’s class- and method- level tests into a single test suite (\textit{Y-Comb} column of Table \ref{tab:llmplain_vs_yate}) leads to much better results. In nearly all projects, the combined coverage exceeds that of either part alone. This suggests that different contextual information (class and method) leads to better testing. 

\begin{tcolorbox}[colback=gray!10, colframe=gray!50,boxrule=0.4pt,arc=1pt,left=6pt,right=6pt,top=4pt,bottom=4pt,fontupper=\small,before skip=8pt,after skip=8pt,enhanced]
\textbf{Finding 4:} While method-level testing generates test suites with higher coverage than class-level testing, the combination of the two consistently yields significantly higher coverage.
\end{tcolorbox}
\begin{table}[t]
\centering
\vspace{-0.5em}
\caption{Statistical comparison of YATE with LLM - Plain}
\vspace{-1.0em}
\resizebox{\linewidth}{!}{%
\begin{tabular}{p{0.5\linewidth} c c c c}
\toprule
\textbf{Comparison} & \textbf{Metric} & \textbf{p-value} & \textbf{Significant} & \textbf{VDA} \\
\midrule
Plain (Class) vs. Plain (Method) & Line & 0.0075 & Yes & 0.451 \\
\textbf{YATE(Class) vs. YATE(Method)} & \textbf{Line} & \textbf{0.874} & \textbf{No} & \textbf{0.497} \\
YATE (Class) vs. Plain (Class) & Line & 1.78e-10 & Yes & 0.616 \\
YATE (Method) vs. Plain (Method) & Line & 9.51e-06 & Yes & 0.581 \\
Plain (Class) vs. Plain (Method) & Branch & 0.0012 & Yes & 0.414 \\
\textbf{YATE(Class) vs. YATE(Method)} & \textbf{Branch} & \textbf{0.263} & \textbf{No} & \textbf{0.470} \\
YATE (Class) vs. Plain (Class) & Branch & 1.67e-13 & Yes & 0.696 \\
YATE (Method) vs. Plain (Method) & Branch & 7.31e-08 & Yes & 0.644 \\
\bottomrule
\end{tabular}
}
\label{tab:comparison-plain}
\vspace{-1.0em}
\end{table}

\subsection{RQ3: Ablation Study}

Table \ref{tab:yate-ablation} records the coverage scores of YATE (Class level) and the respective coverage when each component is removed or replaced by the components of LLM-Plain. Notice that we get similar results when examining the method level but do not present them here due to space constraint. The four components that we replace are:
\begin{enumerate}
    \item Generation component: LLM-Plain prompting to generate tests (no Chain-of-Thoughts).
    \item Compiling fixing component: LLM-Plain feedback-driven fixing with a maximum of five iterations.
    \item Oracle fixing component: LLM-Plain feedback-driven fixing with a maximum of five iterations.
    \item Coverage augmentation: Completely removed.
\end{enumerate}

As shown in Table \ref{tab:yate-ablation}, only when all four components are combined together, YATE achieved its highest coverage. Specifically, YATE achieved the lowest line coverage when the generation component was removed, indicating its role to explore more parts of the code. On the contrary, the absence of the coverage augmentation component had minimal impact on line coverage. However, for branch coverage, its absence led to a more pronounced decline: a reduction of 11.09\% in total, and a decrease of 5.38\% on average. 

\begin{table}[t]
\centering
\vspace{-0.5em}
\caption{Ablation study on YATE components}
\vspace{-1.0em}
\resizebox{\linewidth}{!}{
\begin{tabular}{lcccccc}
\toprule
\textbf{} & \multicolumn{2}{c}{\textbf{Line Coverage}} & \multicolumn{2}{c}{\textbf{Branch Coverage}} & \multicolumn{2}{c}{\textbf{Mutation Score}} \\
\cmidrule(lr){2-3} \cmidrule(lr){4-5} \cmidrule(lr){6-7}
\textbf{Configuration} & \textbf{Total} & \textbf{Avg.} & \textbf{Total} & \textbf{Avg.} & \textbf{Total} & \textbf{Avg.} \\
\midrule
YATE – All Components & \textbf{60.71\%} & \textbf{64.94\%} & \textbf{51.06\%} & \textbf{52.26\%} & \textbf{41.39\%} & \textbf{43.98\%} \\
Plain Test Generation & 44.88\% & 44.62\% & 34.64\% & 37.17\% & 32.40\% & 32.29\% \\
Plain Compilation Fixing & 47.63\% & 49.39\% & 31.76\% & 37.16\% & 30.94\% & 32.07\% \\
Plain Oracle Fixing & 49.33\% & 51.63\% & 34.29\% & 39.18\% & 31.89\% & 33.27\% \\
No Coverage Augmentation & 57.59\% & 61.84\% & 39.97\% & 46.88\% & 38.39\% & 40.53\% \\
\bottomrule
\end{tabular}
}
\label{tab:yate-ablation}
\vspace{-0.5em}
\end{table}

When the compilation fixing component was replaced, YATE scored the lowest branch coverage and mutation score than any of the other variations of the ablation study. Specifically, the absence of this component resulted in a 19.30\% drop in total branch coverage and 15.10\% on average, and a 10.45\% reduction in total mutation score and 11.91\% on average.

\begin{tcolorbox}[colback=gray!10, colframe=gray!50,boxrule=0.4pt,arc=1pt,left=6pt,right=6pt,top=4pt,bottom=4pt,fontupper=\small,before skip=8pt,after skip=8pt,enhanced]
\textbf{Finding 5:} All four components of YATE are important. The generation component is essential for a broader line coverage, the coverage augmentation for branch coverage, whereas the most important component is the compilation fixing.  
\end{tcolorbox}

\subsection{RQ4: Sensitivity to LLMs used}

\begin{figure}[b]
    \centering
    \includegraphics[width=0.95\linewidth]{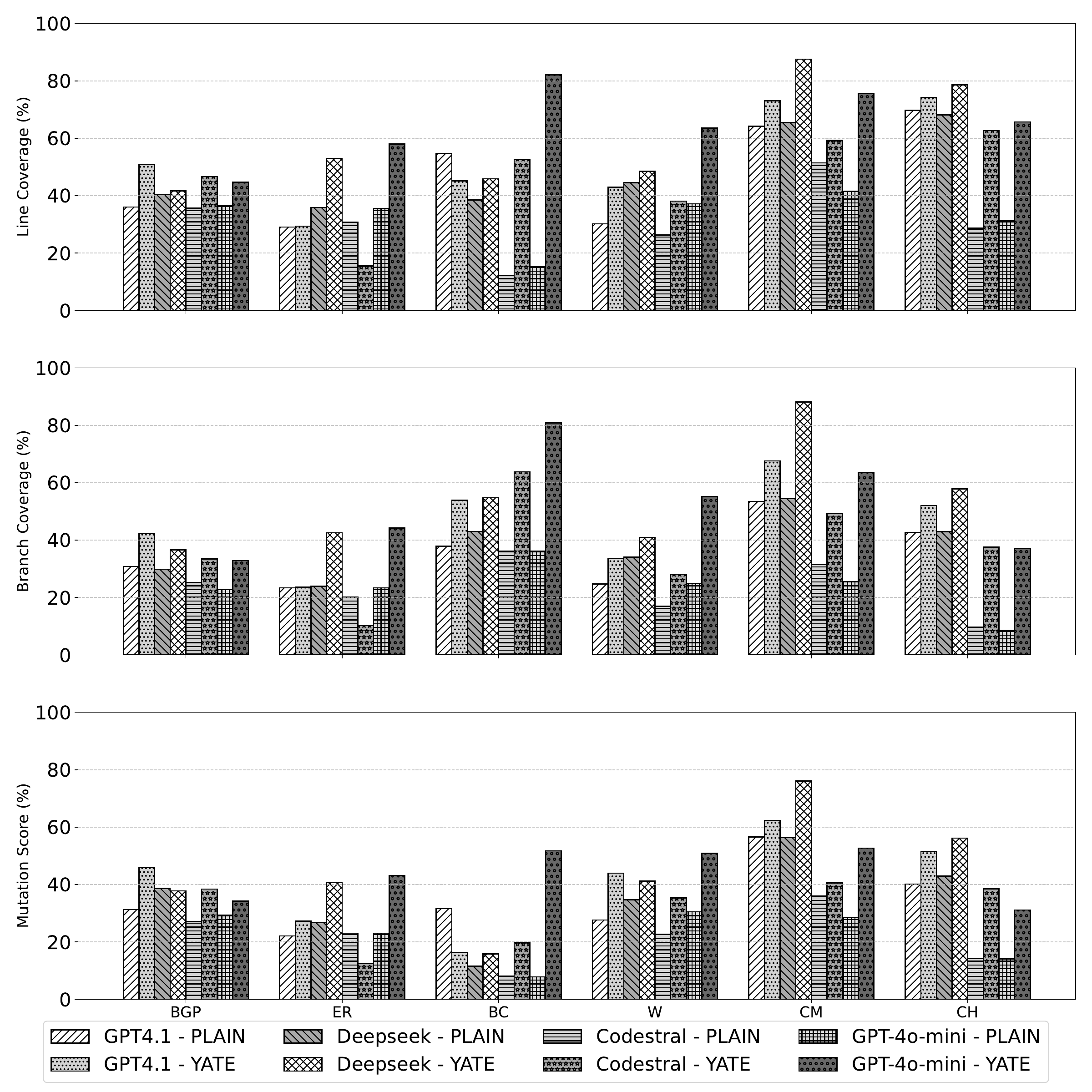}
    \vspace{-1.0em}
    \caption{Plain Vs YATE when using different LLMs.}
    \vspace{-1.0em}
    \label{fig:rq4_results}
\end{figure}

Figure \ref{fig:rq4_results} shows the results of YATE and LLM-Plain when applied with different LLMs. We observe minor differences among the LLMs. Although certain models perform better than others, YATE outperforms LLM-Plain in all but one project, for all models. 

Specifically we observe that YATE outperformed LLM-Plain in 5/6 projects when ran on GPT4.1. LLM-Plain obtained 9.49\% higher line coverage and 15.34\% higher mutation score in \textit{binance-Connector}. We also see that YATE outperformed LLM-Plain in all projects when ran on DeepSeek-V3. The only exception is the mutation score of the project \textit{batch-processing-gateway}, where LLM-Plain killed 0.9\% more mutants. When ran on Codestral, YATE outperformed LLM-Plain in 5/6 projects. On \textit{event-ruler} project LLM-Plain scored better in all metrics.

Table \ref{tab:model-comparison} shows the total coverage and average coverage of 6 studied projects. We can see that when YATE used GPT-4o-mini, it achieved the highest code coverage, surpassing even GPT 4.1 which is a larger and newer model in the GPT family. Followed by GPT-4o-mini, YATE had slightly worse results using Deepseek V3 whereas the results of LLM-Plain improved. Finally, we observe that Codestral, although optimized for code, it scored the worse results for both YATE and LLM-Plain. In all the four models used though, YATE outperformed the LLM-Plain.

\begin{tcolorbox}[colback=gray!10, colframe=gray!50,boxrule=0.4pt,arc=1pt,left=6pt,right=6pt,top=4pt,bottom=4pt,fontupper=\small,before skip=8pt,after skip=8pt,enhanced]
\textbf{Finding 6:} YATE significantly improves over the LLM-Plain method in all models studied.
\end{tcolorbox}

\begin{table}[t]
\centering
\vspace{-0.5em}
\caption{YATE Vs LLM-Plain across different LLMs}
\vspace{-1.0em}
\resizebox{\columnwidth}{!}{
\begin{tabular}{lcccccc}
\toprule
\textbf{} & \multicolumn{2}{c}{\textbf{Line Coverage}} & \multicolumn{2}{c}{\textbf{Branch Coverage}} & \multicolumn{2}{c}{\textbf{Mutation Score}} \\
\cmidrule(lr){2-3} \cmidrule(lr){4-5} \cmidrule(lr){6-7}
\textbf{Model} & \textbf{Total} & \textbf{Avg.} & \textbf{Total} & \textbf{Avg.} & \textbf{Total} & \textbf{Avg.} \\
\midrule
GPT-4o-mini (YATE) & 60.71\% & 64.94\% & 44.27\% & 52.26\% & 41.39\% & 43.98\% \\
GPT-4o-mini (Plain) & 33.53\% & 32.88\% & 21.05\% & 23.58\% & 22.08\% & 22.22\% \\
\midrule
GPT-4.1 (YATE) & 50.75\% & 52.61\% & 38.97\% & 45.49\% & 39.72\% & 41.20\% \\
GPT-4.1 (Plain) & 44.72\% & 47.30\% & 32.09\% & 35.51\% & 32.55\% & 34.91\% \\
\midrule
Deepseek V3 (YATE) & 56.69\% & 59.17\% & 48.71\% & 53.45\% & 44.05\% & 44.65\% \\
Deepseek V3 (Plain) & 47.02\% & 48.80\% & 33.19\% & 38.04\% & 34.41\% & 35.16\% \\
\midrule
Codestral (YATE) & 43.14\% & 45.78\% & 27.32\% & 37.05\% & 28.66\% & 30.83\% \\
Codestral (Plain) & 31.21\% & 30.83\% & 20.35\% & 23.26\% & 21.79\% & 21.88\% \\
\bottomrule
\end{tabular}
}
\label{tab:model-comparison}
\end{table}

\subsection{RQ5: Comparison with Other methods}

\begin{table}[t]
\centering
\vspace{-1.0em}
\caption{Statistical comparison}
\vspace{-1.0em}
\resizebox{\linewidth}{!}{%
\begin{tabular}{p{0.5\linewidth} c c c c}
\toprule
\textbf{Comparison} & \textbf{Metric} & \textbf{p-value} & \textbf{Significant} & \textbf{VDA} \\
\midrule
YATE (Class) vs. Hits & Line & 8.94e-21 & Yes & 0.673 \\
YATE (Class) vs. Symprompt & Line & 1.05e-20 & Yes & 0.672 \\
YATE (Class) vs. Testspark & Line & 4.01e-48 & Yes & 0.768 \\
YATE (Class) vs. Coverup & Line & 1.69e-22 & Yes & 0.680 \\
YATE (Method) vs. Hits & Line & 4.24e-21 & Yes & 0.674 \\
YATE (Method) vs. Symprompt & Line & 5.74e-21 & Yes & 0.673 \\
YATE (Method) vs. Testspark & Line & 1.19e-47 & Yes & 0.766 \\
YATE (Method) vs. Coverup & Line & 9.25e-23 & Yes & 0.681 \\
YATE (Class) vs. Hits & Branch & 1.28e-05 & Yes & 0.616 \\
YATE (Class) vs. Symprompt & Branch & 1.86e-07 & Yes & 0.639 \\
YATE (Class) vs. Testspark & Branch & 2.23e-25 & Yes & 0.775 \\
YATE (Class) vs. Coverup & Branch & 5.41e-10 & Yes & 0.666 \\
YATE (Method) vs. Hits & Branch & 2.35e-07 & Yes & 0.637 \\
YATE (Method) vs. Symprompt & Branch & 1.26e-09 & Yes & 0.662 \\
YATE (Method) vs. Testspark & Branch & 8.05e-28 & Yes & 0.788 \\
YATE (Method) vs. Coverup & Branch & 2.89e-12 & Yes & 0.686 \\
\bottomrule
\end{tabular}
}
\vspace{-1.0em}
\label{tab:comparison-baselines}
\end{table}

\begin{table*}[t]
\centering
\caption{Comparison between YATE and state-of-the-art approaches}
\vspace{-0.5em}
\resizebox{\textwidth}{!}{%
\begin{tabular}{lcccccc|cccccc|cccccc}
\toprule
\textbf{Project} & \multicolumn{6}{c|}{\textbf{Line Coverage}} & \multicolumn{6}{c|}{\textbf{Branch Coverage}} & \multicolumn{6}{c}{\textbf{Mutation Score}} \\
\cmidrule(r){2-7} \cmidrule(r){8-13} \cmidrule(r){14-19}
& HITS & SYMP & TSP & CUP & Y-C & Y-M 
& HITS & SYMP & TSP & CUP & Y-C & Y-M 
& HITS & SYMP & TSP & CUP & Y-C & Y-M \\
\midrule
BPG & 35.5\% & 28.0\% & 17.1\% & 25.2\% & \underline{44.7\%} & \textbf{50.9\%} 
    & \underline{35.9\%} & 30.4\% & 19.0\% & 25.4\% & 32.8\% & \textbf{41.3\%}
    & 27.4\% & 23.8\% & 14.3\% & 19.6\% & \underline{34.3\%} & \textbf{45.7\%} \\
ER  & 9.0\% & 13.2\% & 15.8\% & 10.7\% & \underline{58.0\%} & \textbf{59.0\%}
    & 5.0\% & 8.7\% & 10.1\% & 7.5\% & \underline{44.2\%} & \textbf{46.2\%}
    & 6.6\% & 10.6\% & 11.4\% & 9.1\% & \textbf{43.2\%} & \underline{42.1\%} \\
BC  & 71.5\% & 83.0\% & 18.4\% & \underline{84.1\%} & 82.1\% & \textbf{86.9\%}
    & 79.5\% & 65.8\% & 36.1\% & 70.3\% & \underline{80.8\%} & \textbf{84.5\%}
    & 45.3\% & 49.3\% & 10.8\% & \underline{50.1\%} & \textbf{51.8\%} & 41.1\% \\
W   & 45.3\% & 48.8\% & 18.0\% & 38.7\% & \underline{63.5\%} & \textbf{66.0\%}
    & 41.8\% & 44.0\% & 6.0\% & 31.1\% & \underline{55.2\%} & \textbf{61.6\%}
    & 32.4\% & 39.6\% & 6.8\% & 27.1\% & \underline{50.9\%} & \textbf{58.9\%} \\
CM  & 62.4\% & 54.7\% & 34.6\% & 54.2\% & \underline{75.6\%} & \textbf{82.4\%}
    & 51.5\% & 44.1\% & 27.8\% & 44.8\% & \underline{63.5\%} & \textbf{81.5\%}
    & 44.7\% & 43.0\% & 21.4\% & 43.1\% & \underline{52.7\%} & \textbf{72.7\%} \\
CH  & 65.4\% & 59.5\% & 44.1\% & 60.6\% & \underline{65.7\%} & \textbf{69.8\%}
    & \underline{44.3\%} & 38.3\% & 21.9\% & 39.6\% & 37.0\% & \textbf{47.1\%}
    & \underline{40.6\%} & 36.2\% & 17.6\% & 36.8\% & 31.1\% & \textbf{46.1\%} \\
\midrule
Total & 42.4\% & 41.0\% & 23.7\% & 38.9\% & \underline{60.7\%} & \textbf{64.8\%}
      & 29.5\% & 27.2\% & 16.5\% & 25.1\% & \underline{44.3\%} & \textbf{51.1\%}
      & 28.0\% & 28.4\% & 13.9\% & 26.3\% & \underline{41.6\%} & \textbf{47.9\%} \\
Average & 48.2\% & 47.9\% & 24.7\% & 45.6\% & \underline{64.9\%} & \textbf{69.2\%}
        & 42.9\% & 38.5\% & 20.2\% & 36.4\% & \underline{52.3\%} & \textbf{60.2\%}
        & 32.8\% & 33.7\% & 13.7\% & 31.0\% & \underline{44.0\%} & \textbf{51.1\%} \\

\bottomrule
\end{tabular}%
}
\label{tab:yate-vs-baselines}
\end{table*}

\begin{figure}[t]
    \centering
    \vspace{-1.0em}
    \includegraphics[width=\linewidth]{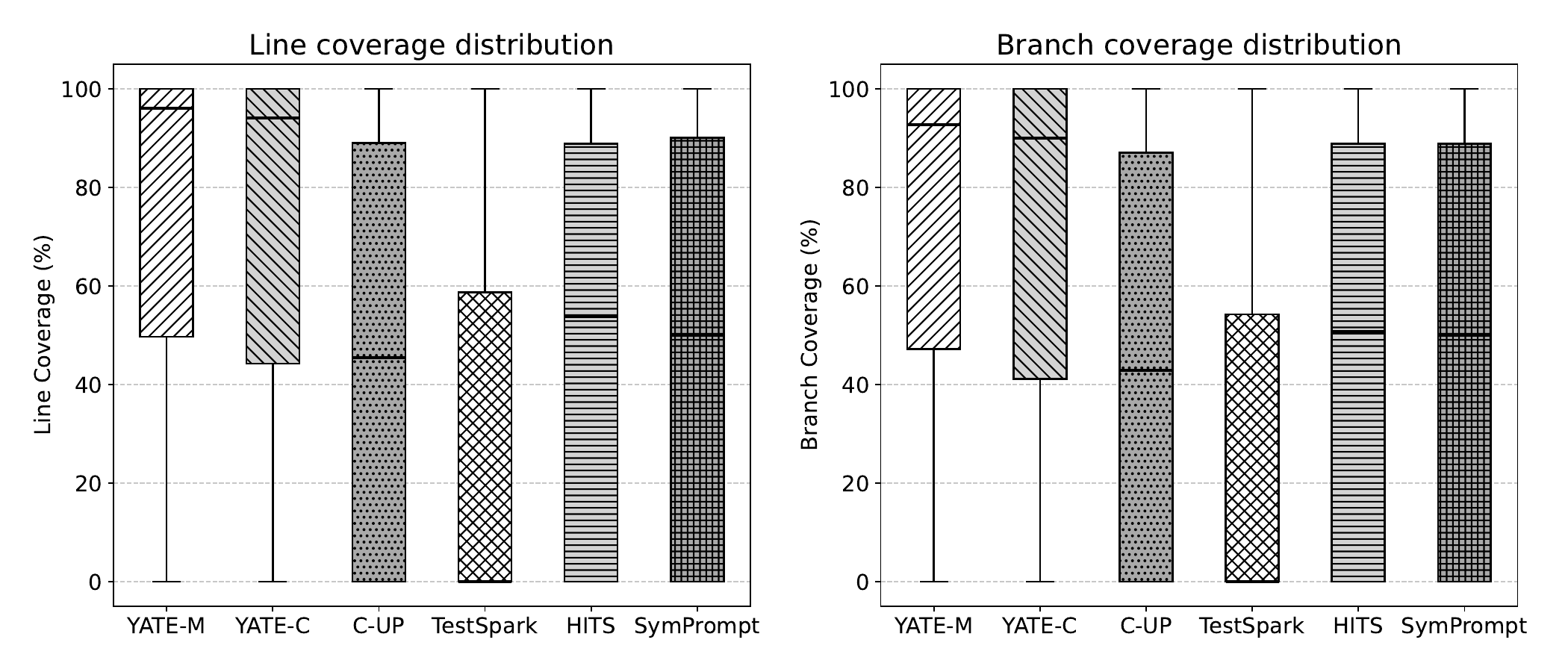}
    \vspace{-2.0em}
    \caption{Distribution of line and branch coverage achieved, per studied class, by the studied methods.}
    \label{fig:coverage_distribution}
    \vspace{-1.0em}
\end{figure}

Table \ref{tab:yate-vs-baselines} records the average coverage achieved by YATE and previous approaches. Figure~\ref{fig:coverage_distribution} shows the distribution of the line and branch coverage achieved, per studied class, by each of the tools. To ensure fair comparison, we have included YATE's method-level results since the baselines also operate at the method level. 
We intentionally exclude results from YATE’s combined test suite, as no existing baseline supports a comparable combination mechanism. In the table, we have highlighted the highest score for each metric and underlined the second-best result. 

We can observe that YATE (method-level) outperformed all prior approaches, in all code coverage metrics and in all projects. The only exception was that YATE (class) outperformed YATE (method) on the mutation score of two repositories. Specifically, for method level testing, YATE achieved on average 64.69\% line coverage which is 37.07\% higher than TestSpark which covered the smaller number of lines. Compared to HITS which had the highest performance after YATE, we see that YATE covered in total 22.36\% more lines. 

A similar trend is observed in both branch coverage and mutation score. While HITS outperforms other baselines, it has significantly less coverage than both of YATE's configurations. Specifically, YATE's method level configuration covered 21.57\% more branches and killed 19.97\% more mutants than HITS. 

Notably, we observe that YATE's class-level configuration also outperformed all previous approaches. It achieved the highest performance in 5 out of 6 repositories for line coverage, 4 out of 6 repositories for branch coverage, and 5 out of 6 repositories for mutation score. In total, YATE (class) covered 18.29\% more lines, 14.80\% more branches, and killed 13.62\% more mutants compared to HITS. Using the Mann–Whitney U test, we found that the performance improvements of YATE over all baseline methods are statistically significant across all metrics and configurations. Table \ref{tab:comparison-baselines} reports the corresponding p-values for each pairwise comparison.

\begin{tcolorbox}[colback=gray!10, colframe=gray!50,boxrule=0.4pt,arc=1pt,left=6pt,right=6pt,top=4pt,bottom=4pt,fontupper=\small,before skip=8pt,after skip=8pt,enhanced]
\textbf{Finding 7:} YATE outperforms prior studies in all evaluation metrics. Its method-level configuration achieves up to 21.57\% higher branch coverage and 19.97\% higher mutation score than the best-performing baseline (HITS). Even the class-level configuration of YATE surpasses all baselines, demonstrating its effectiveness across prior approaches.
\end{tcolorbox}

\begin{table}[ht]
\centering
\caption{Cost (Number of LLM Requests)}
\vspace{-1.0em}
\resizebox{\columnwidth}{!}{%
\begin{tabular}{lrrrrrr}
\toprule
\textbf{Project} & \textbf{HITS} & \textbf{SYMPROMPT} & \textbf{TESTSPARK} & \textbf{COVERUP} & \textbf{YATE (C)} & \textbf{YATE (M)} \\
\midrule
BPG & 3,985 & 994  & \textbf{967}  & 2,047 & 1,032 & 5,154 \\
ER              & 2,530 & 1,081 & 1,100 & 1,579 &  \textbf{755} & 4,980 \\
BC        & 9,167 & 2,693 & 1,404 & 3,545 &  \textbf{476} & 2,794 \\
W                 & 3,216 & 1,381 & 2,189 & 1,916 &  \textbf{789} & 2,412 \\
CM          & 3,703 & 1,493 & 1,597 & 2,222 &  \textbf{816} & 4,035 \\
CH                 & 1,264 &  771 &  706 &  955 &  \textbf{428} & 1,874 \\
\midrule
\textbf{Total}           & 23,865 & 8,413 & 7,963 & 12,264 & \textbf{4,296} & 21,249 \\
\textbf{Average}         & 3,978 & 1,402 & 1,327 & 2,044 & \textbf{716} & 3,542 \\
\bottomrule
\end{tabular}
}
\label{tab:cost}
\end{table}

Table \ref{tab:cost} reports the cost of each approach as a number of LLM queries. Lower values indicate that the approach is less costly. Among all approaches, YATE (class) is consistently the most cost-efficient approach, as it requires 622.27 fewer requests on average than the next best cost-efficient approach, TestSpark. In contrast, YATE (method) and HITS are the two most expensive approaches compared to the other baselines. HITS requires the highest number of requests, with a total of 23,865 requests, averaging 3,977.5 requests per project, which is 436 more than YATE (method) and 3,261.5 more than YATE (class).

These results highlight a trade-off between cost and performance. While YATE (method-level) achieved significantly higher code coverage than all approaches, YATE (class-level) managed to achieve slightly less code coverage with significantly less cost.

\begin{tcolorbox}[colback=gray!10, colframe=gray!50,boxrule=0.4pt,arc=1pt,left=6pt,right=6pt,top=4pt,bottom=4pt,fontupper=\small,before skip=8pt,after skip=8pt,enhanced]
\textbf{Finding 8:} YATE (class-level) is the most cost-efficient approach, requiring significantly less requests yet still outperforming all baselines in code coverage. Although YATE (method-level) delivers the most effective tests, it does so at a higher cost.
\end{tcolorbox}

\section{Discussion}

Interestingly, we found a relatively low performance by previous approaches when compared to the LLM-Plain implementation.
LLM-Plain outperformed all previous approaches in line coverage for 3 out of 6 projects. However, for branch coverage, it achieved the highest coverage in only one project, while HITS was superior in 4 out of 6 projects we study. 

Similarly, SymPrompt achieved the highest branch coverage only in the \textit{Windward} project. When examining the mutation score, LLM-Plain again led in 3 out of 6 projects. We also examined the passing rates of the baselines and the number of generated tests. 

When it comes to passing rates, LLM-Plain had a total passing rate of 53.40\%. Compared to the other baselines, LLM-Plain had the lowest passing rate of all. Hits had a total passing rate of 67.07\%, SymPrompt had a total passing rate of 65.43\%, TestSpark had a total passing rate of 77.19\% and CoverUp 67.47\%. 

LLM-Plain (method) generated a total of 12,158 tests, which is more than 8 times more than TestSpark (1,482 tests), nearly 4.5 times more than SymPrompt (2,719 tests), and 4 times more than CoverUp (3,034 tests). HITS, the baseline with the highest number of generated tests among the state-of-the-art approaches, produced only 8,599 tests, which is 3,559 fewer tests than LLM-Plain. Although the gap in the number of tests that pass is smaller, LLM-Plain still produces more tests. It produced 6,492 passing tests, compared to 5,767 from the second HITS. 

The above results suggest that LLM-Plain is effective because it generates more syntactically valid tests than the other methods. However, LLM-Plain performs worse in branch coverage largely due to the fact that it does not optimise for coverage. 

\section{Threads to Validity}

\textbf{External Validity:} A potential thread to the validity of this study lies to the tools that have been used and the implementations of the algorithms of comparison. To mitigate this threat, we employ one of the state-of-the-art  commercial LLM, GPT 4o-mini, which has been used in previous work~\cite{abs-2503-13580}. To mitigate the threat of tool implementations, we use ChatUniTest, a widely used plugin with the implementation of all algorithms we compare with. It has also been widely used in previous work. ~\cite{WangL0J24}~\cite{abs-2408-03095}~\cite{abs-2501-07425}.  

\textbf{Internal Validity:} YATE's effectiveness is influenced by the capabilities of the underlying model. To mitigate this threat, we evaluated YATE using multiple LLMs with different architectures, including a model specifically optimized for code generation. In addition, YATE's java implementation uses Spoon\cite{PawlakMPNS16} for code analysis and for constructing the required method call graph before the compilation fixing phase. Any bugs, inaccuracies or limitations in Spoon could potentially impact the correctness of our experimental results. To mitigate this threat we manually verified the accuracy of the generated call graphs and ensured that the core analysis steps performed as expected.

\textbf{Construct Validity:} One potential threat to construct validity is data leakage, where the language model may have been exposed to the evaluation data during training. To mitigate this, we selected repositories from prior LLM-based test generation studies~\cite{WangL0J24}~\cite{chatunitest/10.1145/3663529.3663801}, as well as from the publicly available Gitbug-Java dataset~\cite{SilvaSM24}. We manually inspected the creation date of each project to determine whether the primary model of our experiments could have been exposed to them or not. For clarity, we labeled repositories as either seen or unseen based on this analysis.

\section{Conclusion}

In this paper, we studied the role of test repair in the LLM-driven test generation process. Our results showed that focusing on making tests syntactically and semantically valid leads to significant improvements in both the number of generated tests and test effectiveness, leading to substantially higher coverage (more than 20\%) and mutation scores (killing approximately 20\% more mutants) than basic prompting and state-of-the-art methods. We also find that providing LLMs with compilation-relevant feedback, such as class dependencies, to guide them on how to fix their tests, is more important than coverage feedback, suggesting that coverage feedback alone can be misleading on the actual reasons why coverage is not obtained.

\bibliographystyle{ACM-Reference-Format}
\bibliography{main}


\begin{thebibliography}{47}


\ifx \showCODEN    \undefined \def \showCODEN     #1{\unskip}     \fi
\ifx \showDOI      \undefined \def \showDOI       #1{#1}\fi
\ifx \showISBNx    \undefined \def \showISBNx     #1{\unskip}     \fi
\ifx \showISBNxiii \undefined \def \showISBNxiii  #1{\unskip}     \fi
\ifx \showISSN     \undefined \def \showISSN      #1{\unskip}     \fi
\ifx \showLCCN     \undefined \def \showLCCN      #1{\unskip}     \fi
\ifx \shownote     \undefined \def \shownote      #1{#1}          \fi
\ifx \showarticletitle \undefined \def \showarticletitle #1{#1}   \fi
\ifx \showURL      \undefined \def \showURL       {\relax}        \fi
\providecommand\bibfield[2]{#2}
\providecommand\bibinfo[2]{#2}
\providecommand\natexlab[1]{#1}
\providecommand\showeprint[2][]{arXiv:#2}

\bibitem[Alshahwan et~al\mbox{.}(2024a)]%
        {mhetal:fse24-llm}
\bibfield{author}{\bibinfo{person}{Nadia Alshahwan}, \bibinfo{person}{Jubin Chheda}, \bibinfo{person}{Anastasia Finegenova}, \bibinfo{person}{Mark Harman}, \bibinfo{person}{Alexandru Marginean}, \bibinfo{person}{Shubho Sengupta}, {and} \bibinfo{person}{Eddy Wang}.} \bibinfo{year}{2024}\natexlab{a}.
\newblock \showarticletitle{Automated unit test improvement using {Large Language Models} at {Meta}}. In \bibinfo{booktitle}{\emph{{ACM} International Conference on the Foundations of Software Engineering ({FSE 2024})}} (Porto de Galinhas, Brazil, Brazil).
\newblock


\bibitem[Alshahwan et~al\mbox{.}(2024b)]%
        {AlshahwanCFGHHM24}
\bibfield{author}{\bibinfo{person}{Nadia Alshahwan}, \bibinfo{person}{Jubin Chheda}, \bibinfo{person}{Anastasia Finogenova}, \bibinfo{person}{Beliz Gokkaya}, \bibinfo{person}{Mark Harman}, \bibinfo{person}{Inna Harper}, \bibinfo{person}{Alexandru Marginean}, \bibinfo{person}{Shubho Sengupta}, {and} \bibinfo{person}{Eddy Wang}.} \bibinfo{year}{2024}\natexlab{b}.
\newblock \showarticletitle{Automated Unit Test Improvement using Large Language Models at Meta}. In \bibinfo{booktitle}{\emph{{SIGSOFT} {FSE} Companion}}. \bibinfo{publisher}{{ACM}}, \bibinfo{pages}{185--196}.
\newblock


\bibitem[Alshahwan et~al\mbox{.}(2023)]%
        {alshahwan:software}
\bibfield{author}{\bibinfo{person}{Nadia Alshahwan}, \bibinfo{person}{Mark Harman}, {and} \bibinfo{person}{Alexandru Marginean}.} \bibinfo{year}{2023}\natexlab{}.
\newblock \showarticletitle{Software Testing Research Challenges: An Industrial Perspective (keynote paper)}. In \bibinfo{booktitle}{\emph{2023 {IEEE} Conference on Software Testing, Verification and Validation ({ICST 2023})}}. IEEE, \bibinfo{pages}{1--10}.
\newblock


\bibitem[Alshahwan et~al\mbox{.}(2024d)]%
        {mhetal:intense24-keynote}
\bibfield{author}{\bibinfo{person}{Nadia Alshahwan}, \bibinfo{person}{Mark Harman}, \bibinfo{person}{Alexandru Marginean}, \bibinfo{person}{Shubho Sengupta}, {and} \bibinfo{person}{Eddy Wang}.} \bibinfo{year}{2024}\natexlab{d}.
\newblock \showarticletitle{Assured LLM-Based Software Engineering (keynote paper)}. In \bibinfo{booktitle}{\emph{$2^{nd.}$ {ICSE} workshop on Interoperability and Robustness of Neural Software Engineering ({InteNSE})}} (Lisbon, Portugal).
\newblock


\bibitem[Alshahwan et~al\mbox{.}(2024c)]%
        {mhetal:TestGen-obs}
\bibfield{author}{\bibinfo{person}{Nadia Alshahwan}, \bibinfo{person}{Mark Harman}, \bibinfo{person}{Alexandru Marginean}, {and} \bibinfo{person}{Eddy Wang}.} \bibinfo{year}{2024}\natexlab{c}.
\newblock \showarticletitle{Observation-based unit test generation at Meta}. In \bibinfo{booktitle}{\emph{Foundations of Software Engineering ({FSE} 2024)}}.
\newblock


\bibitem[Barr et~al\mbox{.}(2015)]%
        {ebetal:oracle}
\bibfield{author}{\bibinfo{person}{Earl~T. Barr}, \bibinfo{person}{Mark Harman}, \bibinfo{person}{Phil McMinn}, \bibinfo{person}{Muzammil Shahbaz}, {and} \bibinfo{person}{Shin Yoo}.} \bibinfo{year}{2015}\natexlab{}.
\newblock \showarticletitle{The Oracle Problem in Software Testing: {A} Survey}.
\newblock \bibinfo{journal}{\emph{{IEEE} Transactions on Software Engineering}} \bibinfo{volume}{41}, \bibinfo{number}{5} (\bibinfo{date}{May} \bibinfo{year}{2015}), \bibinfo{pages}{507--525}.
\newblock


\bibitem[Bouafif et~al\mbox{.}(2025)]%
        {DBLP:journals/corr/abs-2505-05584}
\bibfield{author}{\bibinfo{person}{Mohamed~Salah Bouafif}, \bibinfo{person}{Mohammad Hamdaqa}, {and} \bibinfo{person}{Edward Zulkoski}.} \bibinfo{year}{2025}\natexlab{}.
\newblock \showarticletitle{{PRIMG} : Efficient LLM-driven Test Generation Using Mutant Prioritization}.
\newblock \bibinfo{journal}{\emph{CoRR}}  \bibinfo{volume}{abs/2505.05584} (\bibinfo{year}{2025}).
\newblock
\urldef\tempurl%
\url{https://doi.org/10.48550/ARXIV.2505.05584}
\showDOI{\tempurl}
\showeprint[arXiv]{2505.05584}


\bibitem[Chen et~al\mbox{.}(2024)]%
        {chatunitest/10.1145/3663529.3663801}
\bibfield{author}{\bibinfo{person}{Yinghao Chen}, \bibinfo{person}{Zehao Hu}, \bibinfo{person}{Chen Zhi}, \bibinfo{person}{Junxiao Han}, \bibinfo{person}{Shuiguang Deng}, {and} \bibinfo{person}{Jianwei Yin}.} \bibinfo{year}{2024}\natexlab{}.
\newblock \showarticletitle{ChatUniTest: A Framework for LLM-Based Test Generation}. In \bibinfo{booktitle}{\emph{Companion Proceedings of the 32nd ACM International Conference on the Foundations of Software Engineering}} (Porto de Galinhas, Brazil) \emph{(\bibinfo{series}{FSE 2024})}. \bibinfo{publisher}{Association for Computing Machinery}, \bibinfo{address}{New York, NY, USA}, \bibinfo{pages}{572–576}.
\newblock
\showISBNx{9798400706585}
\urldef\tempurl%
\url{https://doi.org/10.1145/3663529.3663801}
\showDOI{\tempurl}


\bibitem[Coles et~al\mbox{.}(2016)]%
        {ColesLHPV16}
\bibfield{author}{\bibinfo{person}{Henry Coles}, \bibinfo{person}{Thomas Laurent}, \bibinfo{person}{Christopher Henard}, \bibinfo{person}{Mike Papadakis}, {and} \bibinfo{person}{Anthony Ventresque}.} \bibinfo{year}{2016}\natexlab{}.
\newblock \showarticletitle{{PIT:} a practical mutation testing tool for Java (demo)}. In \bibinfo{booktitle}{\emph{{ISSTA}}}. \bibinfo{publisher}{{ACM}}, \bibinfo{pages}{449--452}.
\newblock


\bibitem[Dakhel et~al\mbox{.}(2024)]%
        {DBLP:journals/infsof/DakhelNMKD24}
\bibfield{author}{\bibinfo{person}{Arghavan~Moradi Dakhel}, \bibinfo{person}{Amin Nikanjam}, \bibinfo{person}{Vahid Majdinasab}, \bibinfo{person}{Foutse Khomh}, {and} \bibinfo{person}{Michel~C. Desmarais}.} \bibinfo{year}{2024}\natexlab{}.
\newblock \showarticletitle{Effective test generation using pre-trained Large Language Models and mutation testing}.
\newblock \bibinfo{journal}{\emph{Inf. Softw. Technol.}}  \bibinfo{volume}{171} (\bibinfo{year}{2024}), \bibinfo{pages}{107468}.
\newblock
\urldef\tempurl%
\url{https://doi.org/10.1016/J.INFSOF.2024.107468}
\showDOI{\tempurl}


\bibitem[DeepSeek{-}AI et~al\mbox{.}(2024)]%
        {abs-2412-19437}
\bibfield{author}{\bibinfo{person}{DeepSeek{-}AI}, \bibinfo{person}{Aixin Liu}, \bibinfo{person}{Bei Feng}, \bibinfo{person}{Bing Xue}, \bibinfo{person}{Bingxuan Wang}, \bibinfo{person}{Bochao Wu}, \bibinfo{person}{Chengda Lu}, \bibinfo{person}{Chenggang Zhao}, \bibinfo{person}{Chengqi Deng}, \bibinfo{person}{Chenyu Zhang}, \bibinfo{person}{Chong Ruan}, \bibinfo{person}{Damai Dai}, \bibinfo{person}{Daya Guo}, \bibinfo{person}{Dejian Yang}, \bibinfo{person}{Deli Chen}, \bibinfo{person}{Dongjie Ji}, \bibinfo{person}{Erhang Li}, \bibinfo{person}{Fangyun Lin}, \bibinfo{person}{Fucong Dai}, \bibinfo{person}{Fuli Luo}, \bibinfo{person}{Guangbo Hao}, \bibinfo{person}{Guanting Chen}, \bibinfo{person}{Guowei Li}, \bibinfo{person}{H. Zhang}, \bibinfo{person}{Han Bao}, \bibinfo{person}{Hanwei Xu}, \bibinfo{person}{Haocheng Wang}, \bibinfo{person}{Haowei Zhang}, \bibinfo{person}{Honghui Ding}, \bibinfo{person}{Huajian Xin}, \bibinfo{person}{Huazuo Gao}, \bibinfo{person}{Hui Li}, \bibinfo{person}{Hui Qu},
  \bibinfo{person}{J.~L. Cai}, \bibinfo{person}{Jian Liang}, \bibinfo{person}{Jianzhong Guo}, \bibinfo{person}{Jiaqi Ni}, \bibinfo{person}{Jiashi Li}, \bibinfo{person}{Jiawei Wang}, \bibinfo{person}{Jin Chen}, \bibinfo{person}{Jingchang Chen}, \bibinfo{person}{Jingyang Yuan}, \bibinfo{person}{Junjie Qiu}, \bibinfo{person}{Junlong Li}, \bibinfo{person}{Junxiao Song}, \bibinfo{person}{Kai Dong}, \bibinfo{person}{Kai Hu}, \bibinfo{person}{Kaige Gao}, \bibinfo{person}{Kang Guan}, \bibinfo{person}{Kexin Huang}, \bibinfo{person}{Kuai Yu}, \bibinfo{person}{Lean Wang}, \bibinfo{person}{Lecong Zhang}, \bibinfo{person}{Lei Xu}, \bibinfo{person}{Leyi Xia}, \bibinfo{person}{Liang Zhao}, \bibinfo{person}{Litong Wang}, \bibinfo{person}{Liyue Zhang}, \bibinfo{person}{Meng Li}, \bibinfo{person}{Miaojun Wang}, \bibinfo{person}{Mingchuan Zhang}, \bibinfo{person}{Minghua Zhang}, \bibinfo{person}{Minghui Tang}, \bibinfo{person}{Mingming Li}, \bibinfo{person}{Ning Tian}, \bibinfo{person}{Panpan Huang}, \bibinfo{person}{Peiyi
  Wang}, \bibinfo{person}{Peng Zhang}, \bibinfo{person}{Qiancheng Wang}, \bibinfo{person}{Qihao Zhu}, \bibinfo{person}{Qinyu Chen}, \bibinfo{person}{Qiushi Du}, \bibinfo{person}{R.~J. Chen}, \bibinfo{person}{R.~L. Jin}, \bibinfo{person}{Ruiqi Ge}, \bibinfo{person}{Ruisong Zhang}, \bibinfo{person}{Ruizhe Pan}, \bibinfo{person}{Runji Wang}, \bibinfo{person}{Runxin Xu}, \bibinfo{person}{Ruoyu Zhang}, \bibinfo{person}{Ruyi Chen}, \bibinfo{person}{S.~S. Li}, \bibinfo{person}{Shanghao Lu}, \bibinfo{person}{Shangyan Zhou}, \bibinfo{person}{Shanhuang Chen}, \bibinfo{person}{Shaoqing Wu}, \bibinfo{person}{Shengfeng Ye}, \bibinfo{person}{Shengfeng Ye}, \bibinfo{person}{Shirong Ma}, \bibinfo{person}{Shiyu Wang}, \bibinfo{person}{Shuang Zhou}, \bibinfo{person}{Shuiping Yu}, \bibinfo{person}{Shunfeng Zhou}, \bibinfo{person}{Shuting Pan}, \bibinfo{person}{T. Wang}, \bibinfo{person}{Tao Yun}, \bibinfo{person}{Tian Pei}, \bibinfo{person}{Tianyu Sun}, \bibinfo{person}{W.~L. Xiao}, {and} \bibinfo{person}{Wangding Zeng}.}
  \bibinfo{year}{2024}\natexlab{}.
\newblock \showarticletitle{DeepSeek-V3 Technical Report}.
\newblock \bibinfo{journal}{\emph{CoRR}}  \bibinfo{volume}{abs/2412.19437} (\bibinfo{year}{2024}).
\newblock


\bibitem[Fan et~al\mbox{.}(2023)]%
        {mhetal:LLM-survey}
\bibfield{author}{\bibinfo{person}{Angela Fan}, \bibinfo{person}{Beliz Gokkaya}, \bibinfo{person}{Mitya Lyubarskiy}, \bibinfo{person}{Mark Harman}, \bibinfo{person}{Shubho Sengupta}, \bibinfo{person}{Shin Yoo}, {and} \bibinfo{person}{Jie Zhang}.} \bibinfo{year}{2023}\natexlab{}.
\newblock \showarticletitle{{L}arge {L}anguage {M}odels for {S}oftware {E}ngineering: {S}urvey and Open Problems}. In \bibinfo{booktitle}{\emph{{ICSE} {F}uture of {S}oftware {E}ngineering ({FoSE} 2023)}}.
\newblock


\bibitem[Foster et~al\mbox{.}(2025a)]%
        {foster:mutation}
\bibfield{author}{\bibinfo{person}{Christopher Foster}, \bibinfo{person}{Abhishek Gulati}, \bibinfo{person}{Mark Harman}, \bibinfo{person}{Inna Harper}, \bibinfo{person}{Ke Mao}, \bibinfo{person}{Jillian Ritchey}, \bibinfo{person}{Herv{\'e} Robert}, {and} \bibinfo{person}{Shubho Sengupta}.} \bibinfo{year}{2025}\natexlab{a}.
\newblock \showarticletitle{Mutation-Guided LLM-based Test Generation at Meta}. In \bibinfo{booktitle}{\emph{2025 {ACM} Conference on Foundations of Software Engineering ({FSE 2025})}}. {ACM}.
\newblock
\newblock
\shownote{Also available as arXiv preprint arXiv:2501.12862}.


\bibitem[Foster et~al\mbox{.}(2025b)]%
        {abs-2501-12862}
\bibfield{author}{\bibinfo{person}{Christopher Foster}, \bibinfo{person}{Abhishek Gulati}, \bibinfo{person}{Mark Harman}, \bibinfo{person}{Inna Harper}, \bibinfo{person}{Ke Mao}, \bibinfo{person}{Jillian Ritchey}, \bibinfo{person}{Herv{\'{e}} Robert}, {and} \bibinfo{person}{Shubho Sengupta}.} \bibinfo{year}{2025}\natexlab{b}.
\newblock \showarticletitle{Mutation-Guided LLM-based Test Generation at Meta}.
\newblock \bibinfo{journal}{\emph{CoRR}}  \bibinfo{volume}{abs/2501.12862} (\bibinfo{year}{2025}).
\newblock


\bibitem[Fraser and Arcuri(2011)]%
        {6004309}
\bibfield{author}{\bibinfo{person}{Gordon Fraser} {and} \bibinfo{person}{Andrea Arcuri}.} \bibinfo{year}{2011}\natexlab{}.
\newblock \showarticletitle{Evolutionary Generation of Whole Test Suites}. In \bibinfo{booktitle}{\emph{2011 11th International Conference on Quality Software}}. \bibinfo{pages}{31--40}.
\newblock
\urldef\tempurl%
\url{https://doi.org/10.1109/QSIC.2011.19}
\showDOI{\tempurl}


\bibitem[Fraser and Zeller(2011)]%
        {FraserZ11}
\bibfield{author}{\bibinfo{person}{Gordon Fraser} {and} \bibinfo{person}{Andreas Zeller}.} \bibinfo{year}{2011}\natexlab{}.
\newblock \showarticletitle{Exploiting Common Object Usage in Test Case Generation}. In \bibinfo{booktitle}{\emph{Fourth {IEEE} International Conference on Software Testing, Verification and Validation, {ICST} 2011, Berlin, Germany, March 21-25, 2011}}. \bibinfo{publisher}{{IEEE} Computer Society}, \bibinfo{pages}{80--89}.
\newblock
\urldef\tempurl%
\url{https://doi.org/10.1109/ICST.2011.53}
\showDOI{\tempurl}


\bibitem[Gao et~al\mbox{.}(2023)]%
        {gao2023retrieval}
\bibfield{author}{\bibinfo{person}{Yunfan Gao}, \bibinfo{person}{Yun Xiong}, \bibinfo{person}{Xinyu Gao}, \bibinfo{person}{Kangxiang Jia}, \bibinfo{person}{Jinliu Pan}, \bibinfo{person}{Yuxi Bi}, \bibinfo{person}{Yixin Dai}, \bibinfo{person}{Jiawei Sun}, \bibinfo{person}{Haofen Wang}, {and} \bibinfo{person}{Haofen Wang}.} \bibinfo{year}{2023}\natexlab{}.
\newblock \showarticletitle{Retrieval-augmented generation for large language models: A survey}.
\newblock \bibinfo{journal}{\emph{arXiv preprint arXiv:2312.10997}} \bibinfo{volume}{2}, \bibinfo{number}{1} (\bibinfo{year}{2023}).
\newblock


\bibitem[Gu et~al\mbox{.}(2024)]%
        {abs-2408-03095}
\bibfield{author}{\bibinfo{person}{Siqi Gu}, \bibinfo{person}{Chunrong Fang}, \bibinfo{person}{Quanjun Zhang}, \bibinfo{person}{Fangyuan Tian}, \bibinfo{person}{Jianyi Zhou}, {and} \bibinfo{person}{Zhenyu Chen}.} \bibinfo{year}{2024}\natexlab{}.
\newblock \showarticletitle{TestART: Improving LLM-based Unit Test via Co-evolution of Automated Generation and Repair Iteration}.
\newblock \bibinfo{journal}{\emph{CoRR}}  \bibinfo{volume}{abs/2408.03095} (\bibinfo{year}{2024}).
\newblock
\urldef\tempurl%
\url{https://doi.org/10.48550/ARXIV.2408.03095}
\showDOI{\tempurl}
\showeprint[arXiv]{2408.03095}


\bibitem[Gu et~al\mbox{.}(2025)]%
        {abs-2503-13580}
\bibfield{author}{\bibinfo{person}{Sijia Gu}, \bibinfo{person}{Noor Nashid}, {and} \bibinfo{person}{Ali Mesbah}.} \bibinfo{year}{2025}\natexlab{}.
\newblock \showarticletitle{{LLM} Test Generation via Iterative Hybrid Program Analysis}.
\newblock \bibinfo{journal}{\emph{CoRR}}  \bibinfo{volume}{abs/2503.13580} (\bibinfo{year}{2025}).
\newblock


\bibitem[Harman et~al\mbox{.}(2025)]%
        {mhpohss:harden}
\bibfield{author}{\bibinfo{person}{Mark Harman}, \bibinfo{person}{Peter O’Hearn}, {and} \bibinfo{person}{Shubho Sengupta}.} \bibinfo{year}{2025}\natexlab{}.
\newblock \showarticletitle{Harden and Catch for Just-in-Time Assured LLM-Based Software Testing: Open Research Challenges (keynote paper)}. In \bibinfo{booktitle}{\emph{2025 {ACM} Conference on Foundations of Software Engineering ({FSE 2025})}}. {ACM}.
\newblock
\newblock
\shownote{Also available as arXiv preprint arXiv:2504.16472}.


\bibitem[Liu et~al\mbox{.}(2025)]%
        {abs-2503-14000}
\bibfield{author}{\bibinfo{person}{Runlin Liu}, \bibinfo{person}{Zhe Zhang}, \bibinfo{person}{Yunge Hu}, \bibinfo{person}{Yuhang Lin}, \bibinfo{person}{Xiang Gao}, {and} \bibinfo{person}{Hailong Sun}.} \bibinfo{year}{2025}\natexlab{}.
\newblock \showarticletitle{LLM-based Unit Test Generation for Dynamically-Typed Programs}.
\newblock \bibinfo{journal}{\emph{CoRR}}  \bibinfo{volume}{abs/2503.14000} (\bibinfo{year}{2025}).
\newblock
\urldef\tempurl%
\url{https://doi.org/10.48550/ARXIV.2503.14000}
\showDOI{\tempurl}
\showeprint[arXiv]{2503.14000}


\bibitem[Mann and Whitney(1947)]%
        {mann1947test}
\bibfield{author}{\bibinfo{person}{Henry~B Mann} {and} \bibinfo{person}{Donald~R Whitney}.} \bibinfo{year}{1947}\natexlab{}.
\newblock \showarticletitle{On a test of whether one of two random variables is stochastically larger than the other}.
\newblock \bibinfo{journal}{\emph{The annals of mathematical statistics}} (\bibinfo{year}{1947}), \bibinfo{pages}{50--60}.
\newblock


\bibitem[{Mistral AI}(2024)]%
        {mistral2024codestral}
\bibfield{author}{\bibinfo{person}{{Mistral AI}}.} \bibinfo{year}{2024}\natexlab{}.
\newblock \bibinfo{title}{Meet Codestral: A powerful and efficient open-weight model for code}.
\newblock \bibinfo{howpublished}{\url{https://mistral.ai/news/codestral}}.
\newblock
\newblock
\shownote{Introduces Codestral, a 22B-parameter open-weight code generation model supporting over 80 programming languages}.


\bibitem[Nan et~al\mbox{.}(2025)]%
        {NanG0025}
\bibfield{author}{\bibinfo{person}{Zifan Nan}, \bibinfo{person}{Zhaoqiang Guo}, \bibinfo{person}{Kui Liu}, {and} \bibinfo{person}{Xin Xia}.} \bibinfo{year}{2025}\natexlab{}.
\newblock \showarticletitle{Test Intention Guided LLM-Based Unit Test Generation}. In \bibinfo{booktitle}{\emph{{ICSE}}}. \bibinfo{publisher}{{IEEE}}, \bibinfo{pages}{1026--1038}.
\newblock


\bibitem[Ni et~al\mbox{.}(2024)]%
        {abs-2406-15743}
\bibfield{author}{\bibinfo{person}{Chao Ni}, \bibinfo{person}{Xiaoya Wang}, \bibinfo{person}{Liushan Chen}, \bibinfo{person}{Dehai Zhao}, \bibinfo{person}{Zhengong Cai}, \bibinfo{person}{Shaohua Wang}, {and} \bibinfo{person}{Xiaohu Yang}.} \bibinfo{year}{2024}\natexlab{}.
\newblock \showarticletitle{CasModaTest: {A} Cascaded and Model-agnostic Self-directed Framework for Unit Test Generation}.
\newblock \bibinfo{journal}{\emph{CoRR}}  \bibinfo{volume}{abs/2406.15743} (\bibinfo{year}{2024}).
\newblock


\bibitem[{OpenAI}(2024)]%
        {openai2024gpt4omini}
\bibfield{author}{\bibinfo{person}{{OpenAI}}.} \bibinfo{year}{2024}\natexlab{}.
\newblock \bibinfo{title}{GPT‑4o-mini: advancing cost‑efficient intelligence}.
\newblock \bibinfo{howpublished}{OpenAI blog post, July 18 2024}.
\newblock
\urldef\tempurl%
\url{https://openai.com/index/gpt-4o-mini-advancing-cost-efficient-intelligence}
\showURL{%
\tempurl}
\newblock
\shownote{Introduces GPT‑4o-mini, a small multimodal model with superior reasoning, coding, and math performance at substantially reduced cost}.


\bibitem[{OpenAI}(2025)]%
        {openai2025gpt41}
\bibfield{author}{\bibinfo{person}{{OpenAI}}.} \bibinfo{year}{2025}\natexlab{}.
\newblock \bibinfo{title}{Introducing GPT-4.1 in the API}.
\newblock \bibinfo{howpublished}{OpenAI Blog}.
\newblock
\urldef\tempurl%
\url{https://openai.com/index/gpt-4-1/}
\showURL{%
\tempurl}
\newblock
\shownote{Announces GPT-4.1 model series with improvements in coding, instruction following, and long‑context capabilities}.


\bibitem[Pan et~al\mbox{.}(2024)]%
        {pan2024aster}
\bibfield{author}{\bibinfo{person}{Rangeet Pan}, \bibinfo{person}{Myeongsoo Kim}, \bibinfo{person}{Rahul Krishna}, \bibinfo{person}{Raju Pavuluri}, {and} \bibinfo{person}{Saurabh Sinha}.} \bibinfo{year}{2024}\natexlab{}.
\newblock \showarticletitle{Aster: Natural and multi-language unit test generation with llms}.
\newblock \bibinfo{journal}{\emph{arXiv preprint arXiv:2409.03093}} (\bibinfo{year}{2024}).
\newblock


\bibitem[Papadakis et~al\mbox{.}(2019)]%
        {PapadakisK00TH19}
\bibfield{author}{\bibinfo{person}{Mike Papadakis}, \bibinfo{person}{Marinos Kintis}, \bibinfo{person}{Jie Zhang}, \bibinfo{person}{Yue Jia}, \bibinfo{person}{Yves~Le Traon}, {and} \bibinfo{person}{Mark Harman}.} \bibinfo{year}{2019}\natexlab{}.
\newblock \showarticletitle{Chapter Six - Mutation Testing Advances: An Analysis and Survey}.
\newblock \bibinfo{journal}{\emph{Adv. Comput.}}  \bibinfo{volume}{112} (\bibinfo{year}{2019}), \bibinfo{pages}{275--378}.
\newblock
\urldef\tempurl%
\url{https://doi.org/10.1016/BS.ADCOM.2018.03.015}
\showDOI{\tempurl}


\bibitem[Pawlak et~al\mbox{.}(2016)]%
        {PawlakMPNS16}
\bibfield{author}{\bibinfo{person}{Renaud Pawlak}, \bibinfo{person}{Martin Monperrus}, \bibinfo{person}{Nicolas Petitprez}, \bibinfo{person}{Carlos Noguera}, {and} \bibinfo{person}{Lionel Seinturier}.} \bibinfo{year}{2016}\natexlab{}.
\newblock \showarticletitle{{SPOON:} {A} library for implementing analyses and transformations of Java source code}.
\newblock \bibinfo{journal}{\emph{Softw. Pract. Exp.}} \bibinfo{volume}{46}, \bibinfo{number}{9} (\bibinfo{year}{2016}), \bibinfo{pages}{1155--1179}.
\newblock


\bibitem[Pizzorno and Berger(2024)]%
        {abs-2403-16218}
\bibfield{author}{\bibinfo{person}{Juan~Altmayer Pizzorno} {and} \bibinfo{person}{Emery~D. Berger}.} \bibinfo{year}{2024}\natexlab{}.
\newblock \showarticletitle{CoverUp: Coverage-Guided LLM-Based Test Generation}.
\newblock \bibinfo{journal}{\emph{CoRR}}  \bibinfo{volume}{abs/2403.16218} (\bibinfo{year}{2024}).
\newblock
\urldef\tempurl%
\url{https://doi.org/10.48550/ARXIV.2403.16218}
\showDOI{\tempurl}
\showeprint[arXiv]{2403.16218}


\bibitem[Ryan et~al\mbox{.}(2024)]%
        {10.1145/3643769}
\bibfield{author}{\bibinfo{person}{Gabriel Ryan}, \bibinfo{person}{Siddhartha Jain}, \bibinfo{person}{Mingyue Shang}, \bibinfo{person}{Shiqi Wang}, \bibinfo{person}{Xiaofei Ma}, \bibinfo{person}{Murali~Krishna Ramanathan}, {and} \bibinfo{person}{Baishakhi Ray}.} \bibinfo{year}{2024}\natexlab{}.
\newblock \showarticletitle{Code-Aware Prompting: A Study of Coverage-Guided Test Generation in Regression Setting using LLM}.
\newblock \bibinfo{journal}{\emph{Proc. ACM Softw. Eng.}} \bibinfo{volume}{1}, \bibinfo{number}{FSE}, Article \bibinfo{articleno}{43} (\bibinfo{date}{July} \bibinfo{year}{2024}), \bibinfo{numpages}{21}~pages.
\newblock
\urldef\tempurl%
\url{https://doi.org/10.1145/3643769}
\showDOI{\tempurl}


\bibitem[Sapozhnikov et~al\mbox{.}(2024)]%
        {sapozhnikov2024testspark}
\bibfield{author}{\bibinfo{person}{Arkadii Sapozhnikov}, \bibinfo{person}{Mitchell Olsthoorn}, \bibinfo{person}{Annibale Panichella}, \bibinfo{person}{Vladimir Kovalenko}, {and} \bibinfo{person}{Pouria Derakhshanfar}.} \bibinfo{year}{2024}\natexlab{}.
\newblock \showarticletitle{TestSpark: IntelliJ IDEA's Ultimate Test Generation Companion}. In \bibinfo{booktitle}{\emph{Proceedings of the 2024 IEEE/ACM 46th International Conference on Software Engineering: Companion Proceedings}}. \bibinfo{pages}{30--34}.
\newblock


\bibitem[Sch{\"{a}}fer et~al\mbox{.}(2024)]%
        {SchaferNET24}
\bibfield{author}{\bibinfo{person}{Max Sch{\"{a}}fer}, \bibinfo{person}{Sarah Nadi}, \bibinfo{person}{Aryaz Eghbali}, {and} \bibinfo{person}{Frank Tip}.} \bibinfo{year}{2024}\natexlab{}.
\newblock \showarticletitle{An Empirical Evaluation of Using Large Language Models for Automated Unit Test Generation}.
\newblock \bibinfo{journal}{\emph{{IEEE} Trans. Software Eng.}} \bibinfo{volume}{50}, \bibinfo{number}{1} (\bibinfo{year}{2024}), \bibinfo{pages}{85--105}.
\newblock


\bibitem[Silva et~al\mbox{.}(2024)]%
        {SilvaSM24}
\bibfield{author}{\bibinfo{person}{Andr{\'{e}} Silva}, \bibinfo{person}{Nuno Saavedra}, {and} \bibinfo{person}{Martin Monperrus}.} \bibinfo{year}{2024}\natexlab{}.
\newblock \showarticletitle{GitBug-Java: {A} Reproducible Benchmark of Recent Java Bugs}. In \bibinfo{booktitle}{\emph{{MSR}}}. \bibinfo{publisher}{{ACM}}, \bibinfo{pages}{118--122}.
\newblock


\bibitem[Straubinger et~al\mbox{.}(2025)]%
        {DBLP:conf/icst/StraubingerKL025}
\bibfield{author}{\bibinfo{person}{Philipp Straubinger}, \bibinfo{person}{Marvin Kreis}, \bibinfo{person}{Stephan Lukasczyk}, {and} \bibinfo{person}{Gordon Fraser}.} \bibinfo{year}{2025}\natexlab{}.
\newblock \showarticletitle{Mutation Testing via Iterative Large Language Model-Driven Scientific Debugging}. In \bibinfo{booktitle}{\emph{{IEEE} International Conference on Software Testing, Verification and Validation, {ICST} 2025 - Workshops, Naples, Italy, March 31 - April 4, 2025}}. \bibinfo{publisher}{{IEEE}}, \bibinfo{pages}{358--367}.
\newblock
\urldef\tempurl%
\url{https://doi.org/10.1109/ICSTW64639.2025.10962485}
\showDOI{\tempurl}


\bibitem[Vargha and Delaney(2000)]%
        {VarghaDelaney2000}
\bibfield{author}{\bibinfo{person}{András Vargha} {and} \bibinfo{person}{Harold~D. Delaney}.} \bibinfo{year}{2000}\natexlab{}.
\newblock \showarticletitle{A Critique and Improvement of the "CL" Common Language Effect Size Statistics of McGraw and Wong}.
\newblock \bibinfo{journal}{\emph{Journal of Educational and Behavioral Statistics}} \bibinfo{volume}{25}, \bibinfo{number}{2} (\bibinfo{year}{2000}), \bibinfo{pages}{101--132}.
\newblock
\showISSN{10769986, 19351054}
\urldef\tempurl%
\url{http://www.jstor.org/stable/1165329}
\showURL{%
\tempurl}


\bibitem[Wang et~al\mbox{.}(2025)]%
        {abs-2506-02954}
\bibfield{author}{\bibinfo{person}{Guancheng Wang}, \bibinfo{person}{Qinghua Xu}, \bibinfo{person}{Lionel~C. Briand}, {and} \bibinfo{person}{Kui Liu}.} \bibinfo{year}{2025}\natexlab{}.
\newblock \showarticletitle{On Mutation-Guided Unit Test Generation}.
\newblock \bibinfo{journal}{\emph{CoRR}}  \bibinfo{volume}{abs/2506.02954} (\bibinfo{year}{2025}).
\newblock


\bibitem[Wang et~al\mbox{.}(2023)]%
        {0002WSLCNCZ23}
\bibfield{author}{\bibinfo{person}{Xuezhi Wang}, \bibinfo{person}{Jason Wei}, \bibinfo{person}{Dale Schuurmans}, \bibinfo{person}{Quoc~V. Le}, \bibinfo{person}{Ed~H. Chi}, \bibinfo{person}{Sharan Narang}, \bibinfo{person}{Aakanksha Chowdhery}, {and} \bibinfo{person}{Denny Zhou}.} \bibinfo{year}{2023}\natexlab{}.
\newblock \showarticletitle{Self-Consistency Improves Chain of Thought Reasoning in Language Models}. In \bibinfo{booktitle}{\emph{The Eleventh International Conference on Learning Representations, {ICLR} 2023, Kigali, Rwanda, May 1-5, 2023}}. \bibinfo{publisher}{OpenReview.net}.
\newblock
\urldef\tempurl%
\url{https://openreview.net/forum?id=1PL1NIMMrw}
\showURL{%
\tempurl}


\bibitem[Wang et~al\mbox{.}(2024)]%
        {WangL0J24}
\bibfield{author}{\bibinfo{person}{Zejun Wang}, \bibinfo{person}{Kaibo Liu}, \bibinfo{person}{Ge Li}, {and} \bibinfo{person}{Zhi Jin}.} \bibinfo{year}{2024}\natexlab{}.
\newblock \showarticletitle{{HITS:} High-coverage LLM-based Unit Test Generation via Method Slicing}. In \bibinfo{booktitle}{\emph{Proceedings of the 39th {IEEE/ACM} International Conference on Automated Software Engineering, {ASE} 2024, Sacramento, CA, USA, October 27 - November 1, 2024}}, \bibfield{editor}{\bibinfo{person}{Vladimir Filkov}, \bibinfo{person}{Baishakhi Ray}, {and} \bibinfo{person}{Minghui Zhou}} (Eds.). \bibinfo{publisher}{{ACM}}, \bibinfo{pages}{1258--1268}.
\newblock
\urldef\tempurl%
\url{https://doi.org/10.1145/3691620.3695501}
\showDOI{\tempurl}


\bibitem[Wei et~al\mbox{.}(2022)]%
        {abs-2201-11903}
\bibfield{author}{\bibinfo{person}{Jason Wei}, \bibinfo{person}{Xuezhi Wang}, \bibinfo{person}{Dale Schuurmans}, \bibinfo{person}{Maarten Bosma}, \bibinfo{person}{Ed~H. Chi}, \bibinfo{person}{Quoc Le}, {and} \bibinfo{person}{Denny Zhou}.} \bibinfo{year}{2022}\natexlab{}.
\newblock \showarticletitle{Chain of Thought Prompting Elicits Reasoning in Large Language Models}.
\newblock \bibinfo{journal}{\emph{CoRR}}  \bibinfo{volume}{abs/2201.11903} (\bibinfo{year}{2022}).
\newblock
\showeprint[arXiv]{2201.11903}
\urldef\tempurl%
\url{https://arxiv.org/abs/2201.11903}
\showURL{%
\tempurl}


\bibitem[Wei(2025)]%
        {11042526}
\bibfield{author}{\bibinfo{person}{Wei Wei}.} \bibinfo{year}{2025}\natexlab{}.
\newblock \showarticletitle{Static Analysis and LLM for Comprehensive Java Unit Test Generation}. In \bibinfo{booktitle}{\emph{2025 8th International Conference on Advanced Electronic Materials, Computers and Software Engineering (AEMCSE)}}. \bibinfo{pages}{87--92}.
\newblock
\urldef\tempurl%
\url{https://doi.org/10.1109/AEMCSE65292.2025.11042526}
\showDOI{\tempurl}


\bibitem[Xu et~al\mbox{.}(2025)]%
        {abs-2506-02943}
\bibfield{author}{\bibinfo{person}{Qinghua Xu}, \bibinfo{person}{Guancheng Wang}, \bibinfo{person}{Lionel~C. Briand}, {and} \bibinfo{person}{Kui Liu}.} \bibinfo{year}{2025}\natexlab{}.
\newblock \showarticletitle{Hallucination to Consensus: Multi-Agent LLMs for End-to-End Test Generation with Accurate Oracles}.
\newblock \bibinfo{journal}{\emph{CoRR}}  \bibinfo{volume}{abs/2506.02943} (\bibinfo{year}{2025}).
\newblock


\bibitem[Yang et~al\mbox{.}(2024)]%
        {abs-2404-04966}
\bibfield{author}{\bibinfo{person}{Chen Yang}, \bibinfo{person}{Junjie Chen}, \bibinfo{person}{Bin Lin}, \bibinfo{person}{Jianyi Zhou}, {and} \bibinfo{person}{Ziqi Wang}.} \bibinfo{year}{2024}\natexlab{}.
\newblock \showarticletitle{Enhancing LLM-based Test Generation for Hard-to-Cover Branches via Program Analysis}.
\newblock \bibinfo{journal}{\emph{CoRR}}  \bibinfo{volume}{abs/2404.04966} (\bibinfo{year}{2024}).
\newblock


\bibitem[Yin et~al\mbox{.}(2025)]%
        {abs-2501-07425}
\bibfield{author}{\bibinfo{person}{Xin Yin}, \bibinfo{person}{Chao Ni}, \bibinfo{person}{Xinrui Li}, \bibinfo{person}{Liushan Chen}, \bibinfo{person}{Guojun Ma}, {and} \bibinfo{person}{Xiaohu Yang}.} \bibinfo{year}{2025}\natexlab{}.
\newblock \showarticletitle{Enhancing LLM's Ability to Generate More Repository-Aware Unit Tests Through Precise Contextual Information Injection}.
\newblock \bibinfo{journal}{\emph{CoRR}}  \bibinfo{volume}{abs/2501.07425} (\bibinfo{year}{2025}).
\newblock


\bibitem[Yuan et~al\mbox{.}(2024)]%
        {chattester/10.1145/3660783}
\bibfield{author}{\bibinfo{person}{Zhiqiang Yuan}, \bibinfo{person}{Mingwei Liu}, \bibinfo{person}{Shiji Ding}, \bibinfo{person}{Kaixin Wang}, \bibinfo{person}{Yixuan Chen}, \bibinfo{person}{Xin Peng}, {and} \bibinfo{person}{Yiling Lou}.} \bibinfo{year}{2024}\natexlab{}.
\newblock \showarticletitle{Evaluating and Improving ChatGPT for Unit Test Generation}.
\newblock \bibinfo{journal}{\emph{Proc. ACM Softw. Eng.}} \bibinfo{volume}{1}, \bibinfo{number}{FSE}, Article \bibinfo{articleno}{76} (\bibinfo{date}{July} \bibinfo{year}{2024}), \bibinfo{numpages}{24}~pages.
\newblock
\urldef\tempurl%
\url{https://doi.org/10.1145/3660783}
\showDOI{\tempurl}


\bibitem[Zhu et~al\mbox{.}(1997)]%
        {ZhuHM97}
\bibfield{author}{\bibinfo{person}{Hong Zhu}, \bibinfo{person}{Patrick A.~V. Hall}, {and} \bibinfo{person}{John H.~R. May}.} \bibinfo{year}{1997}\natexlab{}.
\newblock \showarticletitle{Software Unit Test Coverage and Adequacy}.
\newblock \bibinfo{journal}{\emph{{ACM} Comput. Surv.}} \bibinfo{volume}{29}, \bibinfo{number}{4} (\bibinfo{year}{1997}), \bibinfo{pages}{366--427}.
\newblock
\urldef\tempurl%
\url{https://doi.org/10.1145/267580.267590}
\showDOI{\tempurl}


\end{thebibliography}
\end{document}